\documentclass[
journal=pla,
  manuscript=article-type,
  year=2026,
  volume=XX,
]{cup-journal}

\makeatletter
\def\ps@plain{%
  \renewcommand{\@oddhead}{\hfill\thepage}%
  \renewcommand{\@evenhead}{\thepage\hfill}%
  \renewcommand{\@evenfoot}{}%
  \renewcommand{\@oddfoot}{}%
}
\def\@maketitle{%
  \setlength\parindent{\z@}
  \ifnum\cup@author@cnt<\z@\relax
    \cup@warning{No authors defined: At least one author is required}%
  \fi
  \newpage
  \vspace*{\cup@space@pre@title}%
  {%
    \titlefont
    \titlesize
    \let\@fnsymbol\cup@author@fnsymbol
    \let\footnote@org\footnote
    \let\footnote\cup@title@footnote
    \cup@maketitle@suppinfo \@title
    \cup@title@footnote@check
    \global\cup@footnote@cnt\c@footnote
    \@maketitle@title@hook
    \par
  }%
  \vspace*{\cup@space@post@title}%
  {%
    \authorsize
    \authorfont
    \frenchspacing
    \cup@author@list
    \par
  }%
  \vspace*{\cup@space@post@author}%
  {%
    \affilsize
    \affilfont
    \cup@address@list
    \par
  }%
  \vspace*{\cup@space@post@address}%
  {%
    \emailsize
    \emailfont
    \ifcup@email
      \expandafter\cup@contact@details
      \par
      \vspace*{\cup@space@post@email}%
    \fi
  }%
  {%
    \datesize
    \datefont
    \ifcup@dates
      (Received \cup@recvd; revised \cup@revd; accepted \cup@accptd)%
      \vspace*{\cup@space@post@date}%
    \fi
    \ifcup@copyright
      \bgroup\let\thefootnote\relax
      \footnote{\cup@copyright@notice}%
      \egroup
    \fi
  }%
}
\makeatother

\usepackage{amsmath}
\usepackage[nopatch]{microtype}
\usepackage{booktabs}
\usepackage{tikz}
\usetikzlibrary{positioning}
\usepackage{rotating}

\title{Forecast-Necessary Causal Discovery for Nonlinear Political
Panel Data: Feedback, Functional Form, and the Dynamics of
Democratization}

\author{Michael Coppedge}
\affiliation{Department of Political Science, University of Notre Dame,
Notre Dame, IN, USA}
\alsoaffiliation{Kellogg Institute, University of
Notre Dame, Notre Dame, IN, USA}

\author{Dmitry Zaytsev}
\affiliation{Lucy Family Institute for Data \& Society, University of
Notre Dame, Notre Dame, IN, USA}
\alsoaffiliation{Kellogg Institute, University of
Notre Dame, Notre Dame, IN, USA}

\author{Valentina Kuskova}
\affiliation{Lucy Family Institute for Data \& Society, University of
Notre Dame, Notre Dame, IN, USA}
\email[V. Kuskova]{vkuskova@nd.edu}


\begin{document}

\begin{abstract}
A non-significant coefficient in a dynamic panel model need not imply
the absence of a relationship. It may instead reflect heterogeneous
effects averaged toward zero, reciprocal dynamics overlooked by a
recursive specification, or relationships masked by omitted
correlated covariates. Standard linear estimators cannot distinguish
among these possibilities. We develop an inferential workflow for
political panel data that resolves this ambiguity by combining
flexible autoregressive estimation, forecast-necessity testing,
functional characterization, and same-data linear benchmarking.
Applied to the causal sequence model of democratization on the V-Dem
panel of 113 countries, the workflow reproduces the model's central
finding - the protective belt of civil society, the rule of law, and
institutionalized parties, while recovering reciprocal relationships
from democracy to its institutional supports that a linear model
cannot detect. Three weak published direct effects, two null and one
marginally significant, receive three different diagnoses: one
dissolves under the full specification, one reflects averaged
heterogeneity, and one was masked by the reduced variable set. The
workflow corrects the published record in both directions, removing
one relationship and recovering two, and provides a framework for
evaluating dynamic political theories under a model class that
represents nonlinear and reciprocal mechanisms while preserving
relationship-level interpretation and explicit inferential
standards.
\end{abstract}

\section{Introduction}
\label{sec:intro}

Null results carry a lot of weight in comparative politics. Whether
economic development causes democratization is one of the oldest
quantitative questions in the field \parencite{lipset1959}. After decades of panel evidence, the direct effect of income on democracy remains weak and fragile, and researchers continue to disagree about what such an estimate means and which estimator should produce it. 
\parencite{przeworski2000, boix2003, epstein2006, teorell2010, gassebner2013}.
Development may have no direct effect on democracy; its effect may
be strong in some political contexts and absent in others, averaging
toward zero in a pooled model; or, the effect may run through
variables that the estimated model leaves out. Different evidence is consistent with different theories.

The problem is general. In standard panel estimators, a
non-significant coefficient is consistent with at least four
situations that existing methods are not designed to parse. The relationship may be genuinely absent. It may be
heterogeneous: strong in one region of the data, weak or reversed
in another, so that a constant-coefficient estimator averages it
away \parencite{capoccia2010}. It may be reciprocal: democracy strengthens civil society, which strengthens democracy in turn \parencite{putnam1994, coppedge2022}, and a linear estimator either
excludes the returning arrow by construction or estimates it and
finds nothing. Or, the relationship may run through omitted
variables, because multicollinearity forces the analyst to work with a subset of the theoretically relevant covariates. Vector
autoregressions, structural equation models, and their panel
variants produce the same non-significant estimate in all four situations. When a coefficient is estimated to be around zero, the analyst
cannot say whether the theory has failed, holds only in certain
contexts, or has been obscured by the specification. This article
provides the methodology for telling these situations apart.

Machine learning estimators now make it possible to relax the
restrictions that create the ambiguity, all at once: relationships
can take flexible functional forms \parencite{storm2020machine}, all candidate variables can
enter jointly \parencite{tsochantaridis2004support}, feedback can be estimated rather than ignored \parencite{paxton2011nonrecursive},
and longer lag structures are computationally feasible
\parencite{grimmer2021, cranmer2017}. What flexible estimators have lacked is discipline. They offer no
principled rule for deciding which estimated relationships to
credit, and their advantages have been argued on predictive rather
than inferential grounds. This article supplies the discipline. We
convert the flexibility of machine learning estimation into an
inferential procedure in which every retention decision, every
heterogeneity claim, and every disagreement with a linear baseline
is subjected to an explicit test.

The workflow proceeds in four steps. We begin by replacing the
constant coefficients of a linear autoregression with a neural
additive autoregression \parencite{bussmann2021}, which lets each
dynamic relationship take its own functional form while
preserving the relationship-by-relationship interpretability on
which applied VAR analysis relies. We then ask not whether an
estimated relationship is statistically significant but whether it
is necessary for prediction. A forecast-necessity test
\parencite{diebold1995, flairs2026necessity} checks whether removing
the relationship degrades out-of-sample forecasts, with a retention
threshold determined automatically rather than chosen by the
researcher. For relationships that survive, conditional expectation
analysis \parencite{goldstein2015, icml2026funcval} characterizes
how their direction and magnitude vary across democratic contexts,
revealing whether a weak average effect conceals systematic
heterogeneity. Finally, we estimate a linear benchmark on the
identical data and variable set, so that any disagreement between
the two models can be attributed to estimator flexibility rather
than to specification. Together, these steps turn a single
non-significant coefficient into a diagnosis of a relationship: absent, heterogeneous,
reciprocal, or masked by omission.

We demonstrate the workflow by re-estimating the causal sequence
model of democratization developed by \textcite{coppedge2022} on the
same V-Dem panel of 113 countries ("the baseline model"), expanding the variable set from 10 to 23 (presented in~\ref{app:variables}). Results are centered on three main findings. 

First, the baseline's
central finding survives the removal of every functional restriction
used to establish it: the ``protective belt'' of civil society, rule
of law, and institutionalized parties remains the dominant proximate
determinant of polyarchy, and every belt relationship is retained
under all seeds of a robustness ensemble. Second, estimator
flexibility delivers gains that specification alone cannot: the
flexible model improves out-of-sample forecast accuracy by roughly
ten percent over the best linear competitor. Moreover, it recovers
reciprocal relationships required for that accuracy and stable
across the ensemble, which the linear model misses even with the
full 23-variable specification. 

Third, three weak and contested direct effects in the same
dataset, two of which are null, and one has a small negative
estimate, receive three different diagnoses, and the corrections
run in both directions. The direct effect of agricultural income
dissolves: the baseline reports a marginally significant negative
estimate, but neither estimator retains a direct relationship once
the full variable set enters, and the influence of agrarian
structure operates entirely through intermediate channels. The
direct effect of GDP per capita reflects averaged heterogeneity: the
linear model finds nothing on the identical data, while the flexible
model finds a necessary relationship whose direction varies across
the range of development. The direct effect of literacy reflects the
baseline's specification constraint: once the variable set expands,
even the linear benchmark detects it. To the linear model, these
three situations are indistinguishable; the workflow separates them.

The article makes four methodological contributions. The first is a
forecast-necessity criterion for dynamic relationships, which
replaces the question of coefficient significance with whether a
relationship is required for out-of-sample prediction. It extends
arguments for the disciplined use of predictive performance in
causal political science \parencite{cranmer2017}. The second is a
framework for interpreting retained relationships through
context-conditional, function-valued effects, which distinguishes
systematic heterogeneity from genuinely null relationships. The third is
same-data benchmarking strategy that attributes disagreement between
flexible and linear models to the estimator or to the specification.
The fourth is the general approach for treatment of structurally exogenous
covariates, such as geography, religious composition, and colonial
legacy, which allows them to be entered into the model strictly as causes. Replication code, a seed ensemble, and complete per-run outputs accompany the article.

The proposed workflow is intended to strengthen empirical inference,
not to replace established principles of causal identification. Its
forecast-necessity criterion should therefore not be interpreted as a
claim of counterfactual causation. Rather, the estimated graph
provides a disciplined characterization of a system's dynamic
dependencies and a demanding empirical benchmark that proposed causal
accounts should be able to explain. At the same time, the flexibility
of neural estimation introduces Monte Carlo variation that must itself
be treated as an object of inference rather than ignored. We therefore
quantify estimation stability and report it alongside every
substantive finding. Section~\ref{sec:scope} discusses these
limitations in detail, clarifying both the inferential claims the
workflow supports and those that remain beyond its scope.

The remainder of the article proceeds as follows.
Section~\ref{sec:limits} traces the ambiguity to four restrictions
of standard linear dynamic models. Section~\ref{sec:pipeline}
presents the workflow. Section~\ref{sec:application} describes the
data and research design. Section~\ref{sec:results} reports the
findings. Section~\ref{sec:scope} discusses scope conditions and
limitations, and conclusion is presented in Section~\ref{sec:conclusion}.

\section{The Limits of Linear Dynamic Models in Political Science}
\label{sec:limits}

Dynamic theories in political science are typically evaluated using a
small family of linear dynamic models, most prominently panel vector
autoregressions and structural equation models. These approaches have earned their place in empirical research: they are transparent, can be estimated on the sample sizes common in comparative politics, and produce coefficients with clear substantive interpretations \parencite{freedman2009statistical}. Precisely because of these advantages, however, they rely on a set of structural assumptions that shape the inferences they can support \parencite{lipset1959, capoccia2010}. 

The assumptions become most consequential where dynamic theories make their strongest claims. Four are central to this article. One allows heterogeneous effects to appear as null relationships, another precludes or obscures reciprocal dynamics, a third forces theoretically relevant variables out of the model before estimation,
and a fourth limits the temporal horizon over which dynamic processes
can be represented. Taken together, these assumptions explain why an apparently non-significant relationship can reflect several fundamentally different underlying data-generating processes.

\subsection{Recursivity: One-Way Models for Two-Way Theories}

Many dynamic models in political science, including path-analytic and SEM implementations,  are recursive: causal relationships are specified to flow in one direction, from exogenous variables to endogenous outcomes. This assumption simplifies estimation and interpretation, but it also excludes an important class of theories whose central mechanism is reciprocal
reinforcement. Theories of democratic consolidation, for example,
argue that democracy strengthens civil society even as civil society
helps sustain democracy; institutional quality and democratic
performance are likewise often understood as mutually reinforcing
processes \parencite{putnam1994, przeworski2000, levitsky2018}. Within a recursive
specification, however, only one direction of such a relationship can
be estimated. The reciprocal path is not merely omitted from the
results; it is excluded by construction. The consequence is a
systematic mismatch between theory and empirical test: one of the
defining mechanisms of many dynamic political theories cannot be
represented within the model used to evaluate them
\parencite{blackwell2013}.

\subsection{Linearity: Constant Effects Rule Out Interesting Cases}

Linear dynamic models summarize each relationship with a single
coefficient, implicitly assuming that its effect is constant across
the range of the predictor and across the evolving state of the
system. Yet many political theories make precisely the opposite
claim. Democratization may accelerate only after a critical level of
development is reached, institutional reforms may exhibit
diminishing returns, and the effects of political competition may
strengthen, weaken, or even reverse as regimes evolve.

Linearity in parameters does not preclude nonlinearity in
variables: interactions, polynomial terms, and transformations can
all represent curved and conditional relationships, but each
requires the analyst to specify before estimation which
relationships bend, on which moderator, and in what form. This is a constraint
that limits evaluation in high-dimensional dynamic systems, and
whose failures in routine practice are well documented
\parencite{brambor2006, hainmueller2019, beck2000}. We show the development of this point in~\ref{app:remedies}.

When the underlying relationship bends and the specification does
not, a linear estimator averages fundamentally different local
effects into a single global coefficient. Strong positive effects in
one part of the data and weak, absent, or negative effects elsewhere
can cancel, producing an estimate that is statistically
indistinguishable from zero \parencite{icml2026funcval}. The result is an inferential ambiguity
rather than a simple estimation error: the same non-significant
coefficient is consistent with a genuinely absent relationship, one
whose effects vary systematically across contexts, and one whose
functional form was simply not among those specified. As
Section~\ref{sec:results} demonstrates, one of the best-known null
findings in the democratization literature belongs to the second
category.

\subsection{Dimensionality: Multicollinearity Forces Theory-Driven Omission}

Dynamic political processes are characterized by tightly interconnected
phenomena. Measures of civil society, rule of law, party
institutionalization, state capacity, and other core constructs often
move together, producing substantial multicollinearity. In the panel we
analyze in this study, some pairwise correlations reach 0.9. As the number of
correlated predictors grows, coefficient estimates become increasingly
unstable, and the conventional response is to simplify the model \parencite{kyriazos2023dealing}. The most common simplification is omission: analysts retain the variables judged
most central to the theory and drop the remainder. Omission, however, is not the only remedy: factor scores, reflective latent
variables, and regularized estimation all stabilize inference under
collinearity \parencite{bollen1989, jackman2008, stock2016}, but
aggregation surrenders exactly the relationship-level questions
under study here, and shrinkage does not indicate which retained
relationships are necessary (\ref{app:remedies}).

Whether by omission or by aggregation, then, the analyst chooses
before estimation which distinctions the model will be able to
draw. The baseline democratization model that we revisit
follows this strategy, reducing an initial set of 23 candidate
variables to 10 \parencite{coppedge2022}. Such omissions, however, are not merely technical.
Every excluded variable represents a pathway through which influence may
operate, making an estimated null difficult to distinguish from an
effect that has been displaced into the omitted portion of the system.
Moreover, because the choice of which variables to retain necessarily
precedes estimation, it introduces an additional layer of
researcher-dependent specification. An estimator that remains stable in
the presence of many correlated predictors relaxes this constraint,
allowing theoretical completeness without sacrificing empirical
tractability.

\subsection{Lag Depth: Short Memory for Long-Run Dynamics}

Dynamic political theories are rarely theories of immediate
adjustment: institutional development, state building, and
democratic consolidation unfold over decades \parencite{dark2016waves, pierson2000not}. Applied panel VARs
and structural equation models nonetheless estimate one lag,
occasionally two, because parameters proliferate with lag order,
leaving the model's temporal horizon far shorter than the
theory's \parencite{dormann2015optimal}. A model with only one
or two years of memory, therefore, asks a different empirical question
from the one posed by theories of gradual political development.
Representing longer-term dynamics requires estimators whose complexity
grows much more slowly with lag depth, allowing richer temporal
dependencies to be estimated without sacrificing statistical
stability.

None of these assumptions is a flaw in the linear dynamic toolkit.
Each represents a principled trade-off, exchanging expressive power
for interpretability, statistical efficiency, and tractable
estimation. The argument we make is not that these trade-offs
were misguided, but that they are no longer universally necessary.

\section{A Forecast-Necessity Pipeline for Panel Time Series}
\label{sec:pipeline}

\begin{figure}[t]
\centering
\includegraphics[width=1.0\textwidth]{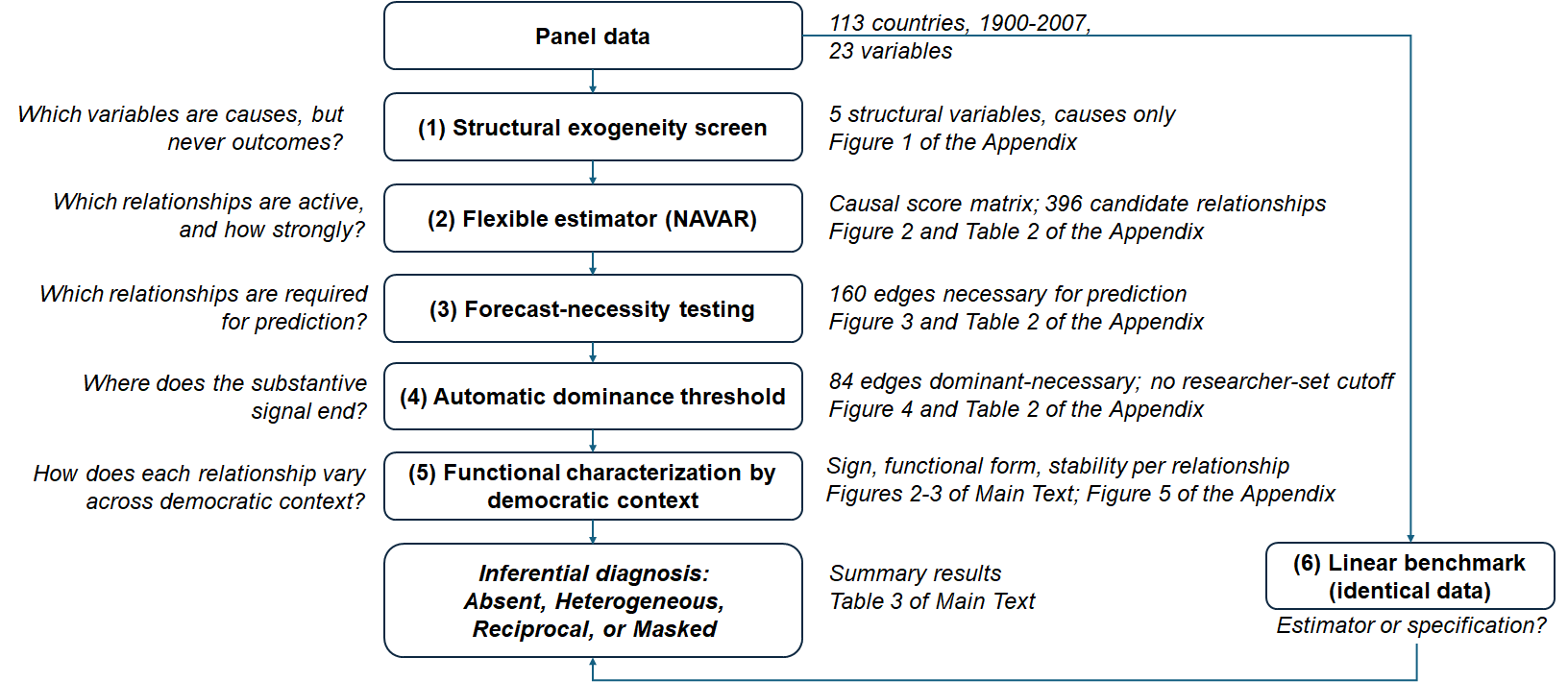}
\caption{The workflow applied to the causal sequence model. The questions at left state what each stage determines, independent
of the application; the quantities at right report each stage's
output for the present one. Each
stage's output for the present application appears at right: the
exogeneity screen admits five structural variables as causes only;
flexible estimation yields a causal score matrix over 396 candidate
relationships; forecast-necessity testing retains 160; the automatic
threshold retains 84; and functional characterization assigns each
retained relationship a sign, a functional form, and a stability
measure across democratic contexts. A linear benchmark, estimated on
identical data and variables, enters at the final stage: the
diagnosis of each relationship follows from the comparison between
the two model classes.}
\label{fig:workflow}
\end{figure}

In this section, we present the workflow that is designed to address the
inferential ambiguities that arise from the assumptions of the
standard linear dynamic toolkit while preserving the properties that
have made it central to empirical political research (relationship-level interpretability, explicit hypothesis testing, and transparent empirical reasoning). Figure~\ref{fig:workflow} summarizes the logic of the workflow and the inferential role of each stage before the individual components are described in detail. Rather than replacing relationship-level interpretation with an opaque predictive model, each stage extends a familiar component of dynamic analysis. 

A preliminary screen (1) first designates structurally exogenous
covariates, which enter every stage as candidate causes but never as
outcomes. Flexible autoregression (2) replaces constant coefficients
with function-valued relationships that can represent heterogeneous
effects. Forecast necessity (3) shifts attention from statistical
significance to whether a relationship contributes to out-of-sample
prediction. An automatic dominance threshold (4) then separates the
relationships that organize the system's dynamics from those whose
contribution is marginal, with the cutoff learned from the
distribution of forecast-loss differentials rather than set by the
researcher. Functional characterization (5) describes how each
retained relationship changes across democratic contexts, revealing
the sources of heterogeneity hidden by a single coefficient.
Finally, a linear benchmark (6) estimated on the identical data
separates the consequences of estimator flexibility from those of
model specification. 

Each component builds on methods validated in prior
work \parencite{flairs2026necessity, icml2026funcval,
kdd2026dcnar}. Here, we emphasize their role in political
inference, while the computational and implementation details are
deferred to in~\ref{app:strexo}-~\ref{app:benchmark}.

\subsection{Structural Exogeneity and Asymmetric Masking}
\label{sec:masking}

Dynamic forecasting models typically treat all variables
symmetrically, allowing each to function both as a predictor and as
an outcome. For an important class of variables in comparative
politics, however, this assumption is neither theoretically nor
substantively appropriate.  The workflow distinguishes
the two roles a variable can play: structural
covariates enter the model as potential causes but are excluded as candidate outcomes. To do so, we rely on well-established procedures \parencite{heck2001multilevel}; details are provided in~\ref{app:strexo}. 

\subsection{Flexible Autoregression with Relationship-Level Structure}
\label{sec:navar}

The foundation of the workflow is a neural additive vector
autoregression \parencite[NAVAR;][]{bussmann2021}, which preserves the
additive structure of a standard VAR while relaxing its assumption of
linear relationships. Importantly, this changes the functional form of individual relationships without changing the object being modeled: each
lagged variable continues to make a distinct contribution to the
prediction of each outcome. 

In a VAR with $K$ lags, the prediction of variable $j$ is a sum of linear terms, one for each lagged covariate. NAVAR preserves the sum but generalizes its elements:
\begin{equation}
\hat{y}_{j,t} \;=\; b_j \;+\; \sum_{i=1}^{N}
\underbrace{f_{ij}\!\left(y_{i,\,t-K:t-1}\right)}_{\substack{\text{contribution of variable } i \\ \text{to the prediction of variable } j}},
\label{eq:navar}
\end{equation}
where each $f_{ij}$ is a small neural network that maps the recent
history of variable $i$ into its contribution to the prediction of
variable $j$. The additive decomposition is what preserves the interpretability of the model. Because each contribution enters separately, the influence of every predictor on every outcome remains an explicit object of inference, much as an individual coefficient does in a linear VAR. The crucial difference is that the contribution is now represented by
a function rather than a constant coefficient, allowing
relationships to strengthen, saturate, or reverse across the range of
the data instead of being averaged into a single global effect.

Beyond relaxing linearity, two additional features of the
specification address the remaining assumptions identified in
Section~\ref{sec:limits}. First, all candidate variables enter
the model jointly. Sparsity is imposed through an $\ell_1$ penalty
on the contribution magnitudes, which shrinks inactive relationships
toward zero and keeps estimation stable in the presence of the
strong cross-correlations that force omission in linear practice.
Second, the model accommodates a lag depth of $K = 8$ years, against
the one or two lags typical of applied dynamic models, bringing the temporal horizon of the estimator more closely into
alignment with the long-run dynamics posited by many political
theories. Estimation pools all countries in the panel, with each
country contributing windows of its own history to a shared set of
contribution functions (estimation details are provided in~\ref{app:estimation}).

The estimated model is summarized by a causal score matrix $S$,
where $S_{ij}$ measures how actively variable $i$ moves the
prediction of variable $j$ across the sample. The score is purely a measure of activity: it conveys neither the direction nor the functional form of a relationship, and by itself it provides no basis for statistical inference. Instead, it serves as a screening device, identifying relationships
that warrant further analysis. The remaining components of the
workflow determine which of these relationships are necessary, how
they operate, and how their effects vary across political contexts.

\subsection{Forecast Necessity and Dominance Necessity as a Retention Criterion}
\label{sec:dm}

A large causal score does not establish that a relationship is
needed: with correlated predictors, a score can reflect variation
that other variables explain equally well. The workflow therefore
asks whether the model forecasts worse without the relationship.
Each candidate is ablated exactly from the fitted model, without
re-estimation, and a Diebold--Mariano test \parencite{diebold1995,
flairs2026necessity} determines whether its removal degrades
accuracy on a fixed evaluation set; relationships that pass are
forecast-necessary. A mixture model fitted to the loss differentials
then separates organizing relationships from marginal ones, with no
researcher-specified parameter entering the decision, and those
above the boundary, termed dominant-necessary, constitute the
estimated graph. The rule is fixed before the results are known.

The forecast-necessity construction, the evaluation set and a disclosure concerning its
overlap with the training span, and the rationale for the empirical
threshold for dominance necessity appear in~\ref{app:fnc}. The complete testing record of all relationships, together with the imperfect
correspondence between score magnitude and forecast necessity that
motivates the test, appears in~\ref{app:edges}.

\subsection{Functional Characterization Across Political Contexts}
\label{sec:ice}

The next stage of the pipeline reconstructs the functional form of the relationships through individual conditional expectation analysis \parencite[ICE;][]{goldstein2015}.
The procedure is simple in outline: the value of a source variable
is swept across its observed range while all other inputs are held
at their observed values, and the model's prediction of the target
is recorded at each point. The resulting curve traces how the
contribution of one variable to another behaves over the data, and
in additive models of the form estimated here, the averaged curve
has a formal interpretation as the function-valued influence of the
source on the target \parencite{icml2026funcval}. The sign,
functional form, and context dependence of each relationship become
empirical quantities to be estimated rather than assumptions imposed
by the model. Forecast necessity identifies relationships;
functional characterization identifies mechanisms.

Because dynamic theories locate heterogeneity in the state of the
political system, the curves are computed separately by democratic
context. Observations are partitioned into three regimes according
to the lagged level of electoral democracy, so that ``low,''
``middle,'' and ``high'' refer throughout to the democratic setting
of a country-year rather than to values of the source variable. The choice of conditioning variable is the analyst's, not the
method's: any theoretically motivated partition, by region or
historical period among others, is estimable by repeating the
computation. We condition on democratic context because the
theories under test locate their heterogeneity there
(Section~\ref{sec:scope} discusses alternatives). Within the partition adopted
here, each relationship receives a classification in each regime:
linear, threshold, saturating, or sign-changing. This classification
transforms heterogeneity from an informal explanation for null
findings into an empirically testable property of a relationship. The same curves determine each relationship's direction: signed
slopes are computed within each regime\footnote{Here we understand
`regimes' not as broad categories of regime types but simply as the
level of democracy that a country possesses at a given time.} and
aggregated, and agreement across all three contexts yields the sign
stability reported in Section~\ref{sec:results}. 

A relationship that a pooled linear model reports as null, but that the estimated functions show to be strong at low levels of democracy and flat at high levels, is not absent; it is saturating, and the pooled estimate has averaged its regions together. Section~\ref{sec:results} shows that one of the democratization literature's most familiar null results has exactly this structure. The classification reduces each curve to a label, but no information is lost: every estimated
curve is retained, the headline relationships are plotted in
~\ref{app:ice}, and the
complete set accompanies the replication materials. The same
appendix develops why named shapes, and not raw curves, are what
make heterogeneity claims testable.

\subsection{A Linear Benchmark on Identical Data}
\label{sec:benchmark-method}

Because the workflow changes the estimator and the variable set at
once, any disagreement with an established linear result is
ambiguous on its own terms. The final component therefore estimates
a linear model under conditions identical to those of the flexible
one: the same variables, lag depth, standardization, and evaluation
observations. Under a fixed specification, any differences between the
two models are attributable to the functional form alone. The three-way
comparison of baseline, benchmark, and flexible model then assigns
each disagreement its source: a relationship neither model detects
is genuinely absent from the estimable system; one the flexible
model retains but the linear model rejects owes its recovery to
functional form; one both models detect but the original smaller
specification missed owes its recovery to the variable set. The
benchmark also disciplines the workflow itself. The flexible model
must forecast the system better than its linear competitors, and the
margin is reported rather than presumed, with the test construction
and competitors described in ~\ref{app:benchmark}. Together, the four components turn an
ambiguous estimate into an attributed one; the next section applies
the workflow to the causal sequence model of democratization.

\section{Application: The Causal Sequence Model of Democratization}
\label{sec:application}

\begin{table}[t]
\centering
\caption{Design of the baseline model and of the present
re-estimation. The data, outcome, and conceptual organization of
the variables are held constant; the estimator, variable set, lag
depth, and retention criterion change.}
\label{tab:design}
\small
\begin{tabular}{@{}p{0.24\textwidth}p{0.34\textwidth}p{0.34\textwidth}@{}}
\toprule
 & Baseline \parencite{coppedge2022} & This article \\
\midrule
Data & V-Dem panel, 113 countries, 1900--2001 & Same, standardized \\
Outcome & Electoral Democracy Index & Same \\
Variables & 10 of 23 candidates & All 23 consistently observed \\
Estimator & Linear recursive path model & Neural additive
autoregression \\
Functional form & Constant coefficients & Flexible per relationship \\
Reciprocal effects & Lagged cross-paths; few significant & Estimated
jointly; tested for necessity \\
Lag depth & One to two years & Eight years \\
Retention criterion & Coefficient significance & Forecast necessity
with automatic threshold \\
Heterogeneity & Single global effect & Characterized by democratic
context \\
\bottomrule
\end{tabular}
\end{table}

Methodological advances are most convincing when they are evaluated
against methods that already perform well on problems of substantive
importance, and the causal sequence model of democratization
provides such a benchmark: its ``protective belt'' finding has
become a reference point for subsequent research
\parencite{coppedge2022, cuzan2025}. Table~\ref{tab:design}
summarizes the design of the baseline and of the present
re-estimation. 

The data, outcome, and conceptual organization of
the variables are held constant, so differences between our
estimates and the published results are attributable to
methodological choices rather than to measurement or scope. The
baseline's authors are explicit about its constraints: reciprocal
effects entered only as lagged cross-paths, of which few reached
significance, and 10 of 23 candidate variables were retained, some
exclusions forced by multicollinearity and others made on
theoretical grounds. The application therefore asks whether the
sparse feedback and weak direct effects of the published model
reflect the system or the specification. Panel construction appears
and the complete variable list are presented in \ref{app:variables}.

The screen of Section~\ref{sec:masking} designates five structural
variables: Protestant population share, European-descended
population share, distance to a natural harbor, mean temperature,
and ethnic fractionalization. Their intraclass correlations lie
between 0.987 and 1.000 against a next-highest 0.856, so the 0.88
threshold falls within a clear empirical gap, and the five enter
the analysis exclusively as candidate causes. Three research questions organize the comparison with the baseline,
each posing a progressively more demanding test of the proposed
workflow.

\emph{RQ1 (Replication).} Does the protective belt survive when the
functional restrictions of the linear model are relaxed? Before a new
method can reveal previously unobservable structure, it must first
recover what is already well established. If the protective belt were
an artifact of linearity or recursivity, relaxing those assumptions
should weaken or eliminate it. Conversely, a workflow that cannot
recover known structure provides little basis for interpreting novel
findings.

\emph{RQ2 (Representation).} Which relationships become estimable
once the assumptions of the linear model are relaxed? Guided by the
limitations identified in Section~\ref{sec:limits}, we focus on two
classes of relationships: reciprocal effects running from polyarchy
back to its institutional supports, and relationships involving the
13 variables omitted from the original specification because of
multicollinearity.

\emph{RQ3 (Diagnosis).} When the baseline reports a weak or null direct effect, what does that estimate actually represent? The workflow asks whether each null reflects a genuinely absent relationship, context-dependent heterogeneity, or masking by the
reduced specification. The third question carries the central methodological
argument of the article.

\section{Results}
\label{sec:results}

\begin{table}[t!]
\centering
\caption{The estimated system at a glance. All quantities are from
the seed-42 reference run; ensemble ranges appear in
~\ref{app:ensemble}, and the complete
forecast comparison in ~\ref{app:benchmark}.}
\label{tab:system}
\small
\begin{tabular}{@{}p{0.85\textwidth}r@{}}
\toprule
Quantity & Value \\
\midrule
Candidate relationships (23 candidate causes $\times$ 18 dynamic
outcomes, self-effects excluded) & 396 \\
Forecast-necessary relationships (Diebold--Mariano test) & 160 \\
Dominant-necessary relationships (automatic threshold,
$5.48 \times 10^{-4}$): the estimated graph & 84 \\
\addlinespace
Positively / negatively signed net effects & 47 / 37 \\
Relationships with a stable sign across democratic contexts & 84 of 84 \\
Relationships nonlinear in at least one democratic regime & 73 of 84 \\
\addlinespace
Held-out forecast error (MSE), flexible model & 0.203 \\
Held-out forecast error (MSE), best linear benchmark & 0.227 \\
\bottomrule
\end{tabular}
\end{table}

\begin{figure}[ht!]
\centering
\includegraphics[width=\textwidth]{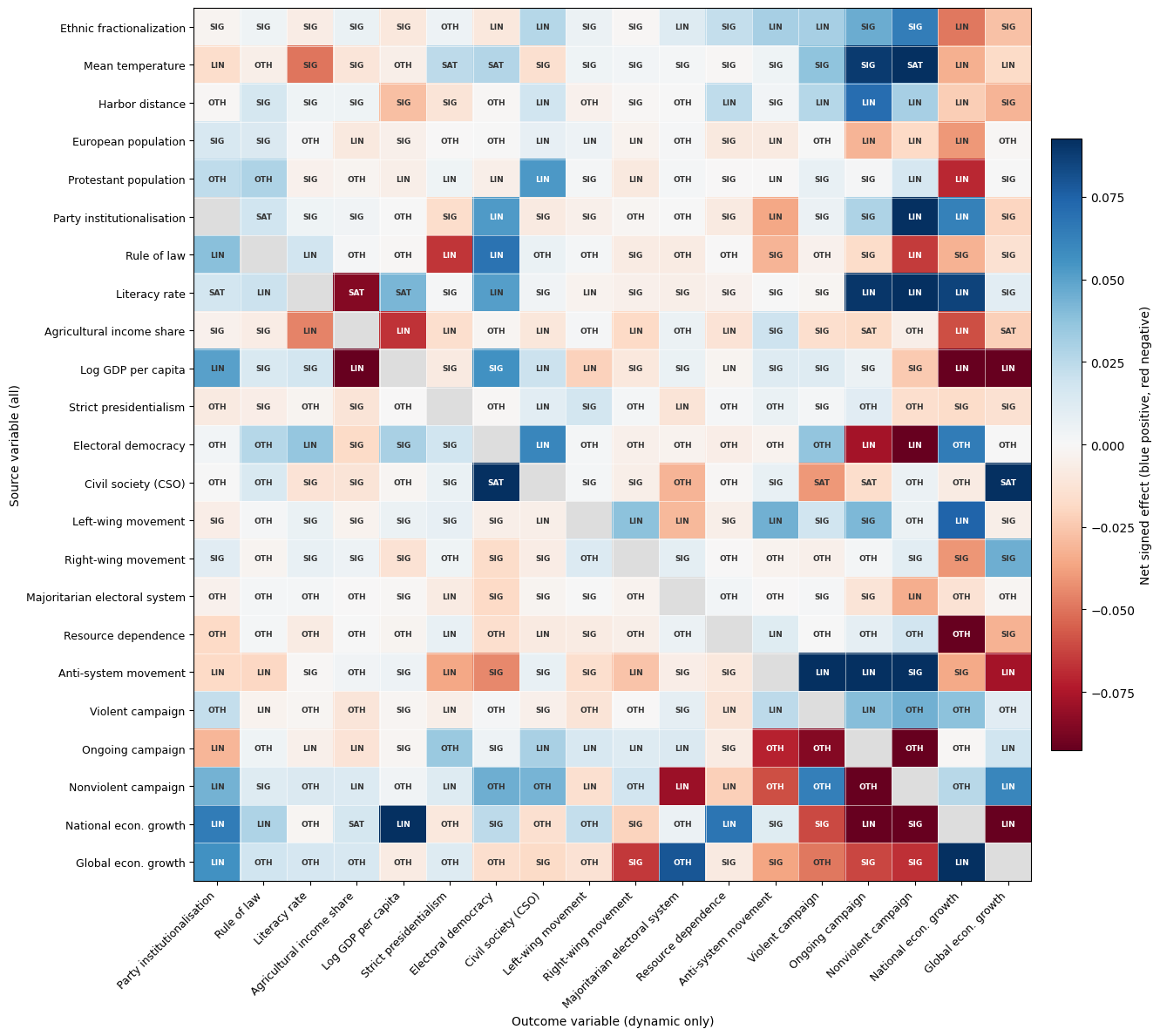}
\caption{Signed net effects and functional forms across the
estimated system. Cell color reports the direction and magnitude of
each source variable's net effect on an outcome, aggregated from the
regime-specific response curves: blue denotes positive effects, red
negative, and grey cells are uncharacterized. Cell text reports the
consensus functional form across democratic contexts (LIN linear,
THR threshold, SAT saturating, SIG sign-changing, OTH other); where
forms differ across regimes, the modal form is shown, with the
per-regime record in Table~\ref{tab:comparison} and in
~\ref{app:ice}. Two
system-level patterns are visible at a glance: the uniformly
positive row for polyarchy, summarizing the reciprocal dynamics of
Section~\ref{sec:rq2}, and the uniformly negative row for
agricultural income, summarizing the indirect suppression
established in Section~\ref{sec:rq3}.}
\label{fig:signedmatrix}
\end{figure}

Applying the workflow to the causal sequence model yields a sparse
but richly nonlinear dynamic structure, summarized in
Table~\ref{tab:system}; all estimates come from a robustness
ensemble of five runs. The analysis begins from the estimated
causal score matrix over the 396 candidate relationships in the
system. Of these, 160 are found to be forecast-necessary, and 84
remain after application of the automatic dominance criterion;
the matrix itself, the relationship-level testing record, and the
fitted mixture that places the threshold appear in
~\ref{app:edges} of the Supplementary Material. These
dominant-necessary relationships constitute the estimated graph.
Despite the prevalence of nonlinear functional forms, every
retained relationship preserves a stable direction across
democratic contexts. Figure~\ref{fig:signedmatrix} displays these
directions together with each relationship's functional form, and
the combination is the system's summary in a single view: sign,
strength, and shape for every characterized relationship.

The composition of the graph points to the same substantive
conclusion. Seventy-three of the 84 retained relationships are
nonlinear in at least one democratic regime, exhibiting threshold,
saturating, or sign-changing behavior. Nonlinearity is therefore not
an occasional refinement of an otherwise linear system but its
dominant empirical characteristic, visible directly in
Figure~\ref{fig:signedmatrix}, where linear cells are the exception
in nearly every row. A modeling framework restricted to constant
effects necessarily fails to represent much of the structure present
in the data.

Out-of-sample forecasting performance reinforces the same
conclusion. On the identical data and variable set,
the flexible model improves held-out forecast accuracy by
approximately ten percent over the best linear benchmark
(Table~\ref{tab:system}; the complete comparison appears in
~\ref{app:benchmark}). The improvement
matters inferentially as well as predictively: forecast necessity is
defined relative to the forecasting model, and the criterion carries
force because the flexible class predicts the system more
accurately.

\subsection{Replicating the Protective Belt}
\label{sec:rq1}

The baseline model's central finding is reproduced after relaxing
the functional assumptions of the linear-recursive specification.
Six relationships leading into polyarchy are retained in the
estimated graph, all positively signed, with the three strongest
corresponding to the protective belt. Civil society remains the
dominant influence, with a causal score 0.073, approximately
twice that of any other predictor, followed by the rule of law
(0.039) and party institutionalization (0.033). All three
relationships are retained under every seed of the robustness
ensemble, with causal scores varying by less than 0.002 across runs
(\ref{app:ensemble}). A structure originally
estimated under a linear, recursive model therefore re-emerges with
the same ordering of influences in a substantially more flexible
modeling framework.

The flexible specification also refines the original result. Rather
than remaining constant across political settings, each of the
three protective-belt relationships is approximately linear in
low-democracy contexts but becomes saturating as democracy reaches
middle and high levels. Civil society, the rule of law, and
institutionalized parties therefore exert their strongest marginal
influence where democratic institutions are weakest, with additional
improvements yielding progressively smaller gains as democracy
consolidates. Because the democracy indices are bounded and
constructed through a transformation that compresses variation near
the top of the scale, some portion of the estimated flattening
could in principle reflect measurement rather than substance;
Section~\ref{sec:scope} weighs this possibility against the
heterogeneity of estimated shapes. It is a pattern that casts the belt as a stabilizing rather
than an accelerating force. This reading is consistent with the
homeostatic account of democratic stability that motivated the
original causal sequence model \parencite{coppedge2022,
capoccia2010}. Replication is thus not the endpoint of the analysis
but its foundation: the workflow not only recovers the established
structure but also identifies the conditions under which that
structure operates.

\subsection{Feedback Becomes Estimable and Proves Necessary}
\label{sec:rq2}

The most important departure from the baseline concerns relationships
that its recursive specification could not represent. In the estimated
graph, polyarchy is not only an outcome but an active source of
institutional change. Nine outgoing relationships from polyarchy are
retained at the reference seed, eight of which are reproduced across
every seed of the robustness ensemble. Democracy feeds back positively
on the institutional conditions associated with its own persistence,
strengthening civil society (0.047), literacy (0.023), and the rule of
law (0.022), and each of these reciprocal relationships proves
necessary for out-of-sample prediction. What appears as a
one-directional causal ordering in the baseline emerges instead as a
system of mutually reinforcing dynamics, providing direct empirical
support for a central claim of democratic consolidation theory
\parencite{putnam1994, levitsky2018}.

The linear benchmark demonstrates that representing reciprocal
relationships is not the same as detecting them. Estimated on the
identical data and variable set, the linear model is free to include
feedback effects but fails to recover them. The estimated effects of
polyarchy on civil society and the rule of law are statistically
indistinguishable from zero under the robust test, while the effect on
literacy is estimated with the opposite sign (\ref{app:benchmark}). Functional characterization explains this
discrepancy. The reciprocal effects on civil society and on literacy
are approximately linear in low-democracy contexts but become
saturating as democracy deepens, mirroring the functional form of the
protective belt itself, and a single coefficient averages such
regime-dependent effects toward zero. The effect on the rule of law
varies in form across regimes without a single dominant pattern
(\ref{app:ice}), a heterogeneity that a
constant coefficient is equally unable to represent. Feedback is
therefore inaccessible to the linear toolkit for two independent
reasons: it is excluded by construction in recursive specifications,
and even when recursive restrictions are removed, its regime-dependent
forms are obscured by constant coefficients.

Relaxing the dimensionality constraint also reveals relationships that
the baseline could not estimate. Among the 13 variables excluded from
the original specification, the most notable new finding concerns
nonviolent campaigns, which exert a positive influence on polyarchy
(causal score 0.030). The estimated response is itself nonlinear:
country-years without active mobilization exhibit essentially no
effect, whereas democratic gains increase approximately linearly once
nonviolent campaigns are under way, consistent with the mechanism
proposed in the civil resistance literature
\parencite{chenoweth2011}. This relationship is less stable than the
preceding findings, appearing in three of the five ensemble runs.
Accordingly, we interpret it as a promising but provisional result
rather than as an established feature of the estimated system.

\subsection{Three Estimates, Three Diagnoses}
\label{sec:rq3}

\begin{figure}[t]
\centering
\includegraphics[width=\textwidth]{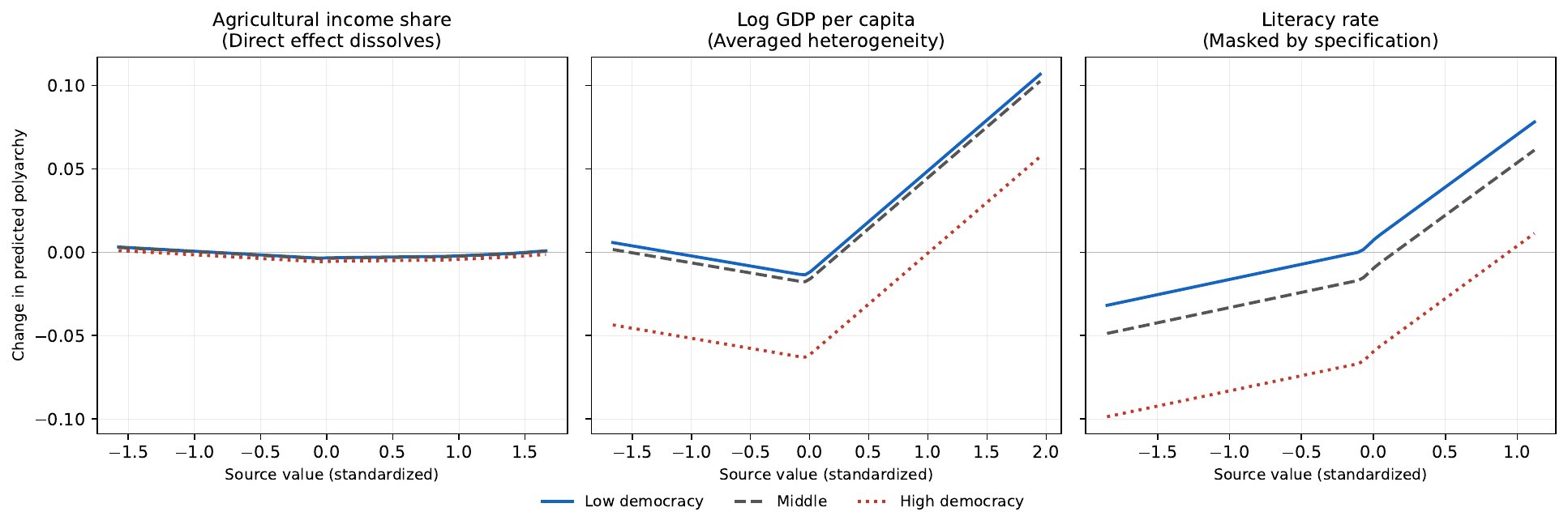}
\caption{Three weak direct effects, three diagnoses. Each panel
shows the estimated response of polyarchy to one variable, computed
separately within low, middle, and high democratic contexts. The
direct effect of agricultural income (left) is flat throughout: no
direct relationship survives once intermediate channels enter the
model, and the baseline's small negative estimate dissolves. The
effect of GDP per capita (center) reverses direction across the
range of development within every context, the pattern that a
constant coefficient averages toward zero. The effect of literacy
(right) is positive throughout and was masked in the baseline by the
reduced variable set: once the specification expands, even a linear
model detects it. In the published model these three direct effects
appear as marginally significant, null, and null; the response
curves show one effect dissolving and two emerging, a correction
running in both directions.}
\label{fig:threenulls}
\end{figure}

The baseline model reports three weak direct effects on polyarchy:
null estimates for literacy and GDP per capita, and a small negative
estimate for agricultural income. Nothing in the published analysis
indicates that these three estimates are weak for different reasons,
and nothing in it could. The workflow shows that they are, and that
correcting the record runs in both directions. Two of the three
effects prove stronger than published, for two different reasons;
the third proves weaker, dissolving entirely once the full system is
estimated. The three cases correspond to three of the inferential
situations identified in Section~\ref{sec:limits}.

The direct effect of agricultural income dissolves (Figure~\ref{fig:threenulls}, left panel). The baseline reports a marginally significant negative estimate, but no direct relationship to polyarchy is retained under any seed of the robustness ensemble, and the linear benchmark, estimated on the full 23-variable specification, likewise detects none. The
variable is nevertheless an active part of the system: six outgoing
relationships are retained, all negative, linking agricultural income
to literacy (0.027), GDP per capita (0.039), campaign activity, and
strict presidentialism. In the estimated system, agrarian economic
structure influences democratization entirely through these
intermediate pathways rather than through a direct effect on
polyarchy \citep{inglehart2005}. The workflow therefore confirms the baseline's mediation 
account while placing it on a stronger inferential foundation based
on explicit relationship-level testing rather than on recursive
modeling assumptions.

The null for GDP per capita reflects averaged heterogeneity rather
than genuine absence. Even when estimated on the identical data and
the full 23-variable specification, the linear benchmark reports no
direct effect ($p = 0.59$ under the robust test). The flexible model,
by contrast, identifies the relationship as forecast-necessary
(causal score 0.033, retained in four of five ensemble runs).
Functional characterization explains the apparent contradiction. The direction of the effect reverses across the range of development itself, negative at low levels of GDP per capita and positive at higher levels, within every democratic context, so that a single linear coefficient averages the opposing segments toward zero.  The crux of the debate about per capita GDP is therefore not, on
this evidence, about an absence of influence but the consequence of forcing
a heterogeneous relationship into a constant-effect specification - a
possibility long recognized in the literature but not previously
amenable to direct empirical evaluation
\parencite{przeworski2000, boix2003, casper2003, epstein2006,
teorell2010, gassebner2013, knutsen2022}. 

The null for literacy reflects the reduced specification of the
baseline model. Once the variable set expands from 10 to 23
predictors, even the linear benchmark identifies a positive direct
effect ($p = 0.002$ under the robust test), indicating that the
original null resulted from omitted structure rather than from the
functional form of the estimator. The flexible model then provides
additional information that the linear benchmark cannot: the
relationship is forecast-necessary (causal score 0.040, retained in
four of five ensemble runs), and its estimated response is
approximately linear in low-democracy contexts before becoming
saturating as democracy deepens. Literacy therefore joins the
protective belt as a stabilizing influence whose marginal effect is
greatest where democratic institutions remain weakest.

Three weak estimates thus receive three distinct diagnoses: one
effect dissolves under the full specification, one was averaged
toward zero by a constant coefficient, and one was masked by the
reduced variable set. The workflow subtracts one relationship from
the published record and adds two - a correction running in both
directions that no single estimator, linear or flexible, could issue
in isolation. The distinction emerges only by combining flexible
estimation, forecast-necessity testing, functional characterization,
and same-data linear benchmarking within a single inferential
workflow.

\subsection{The Estimated System}
\label{sec:network}

\begin{figure}[t]
\centering
\includegraphics[width=\textwidth]{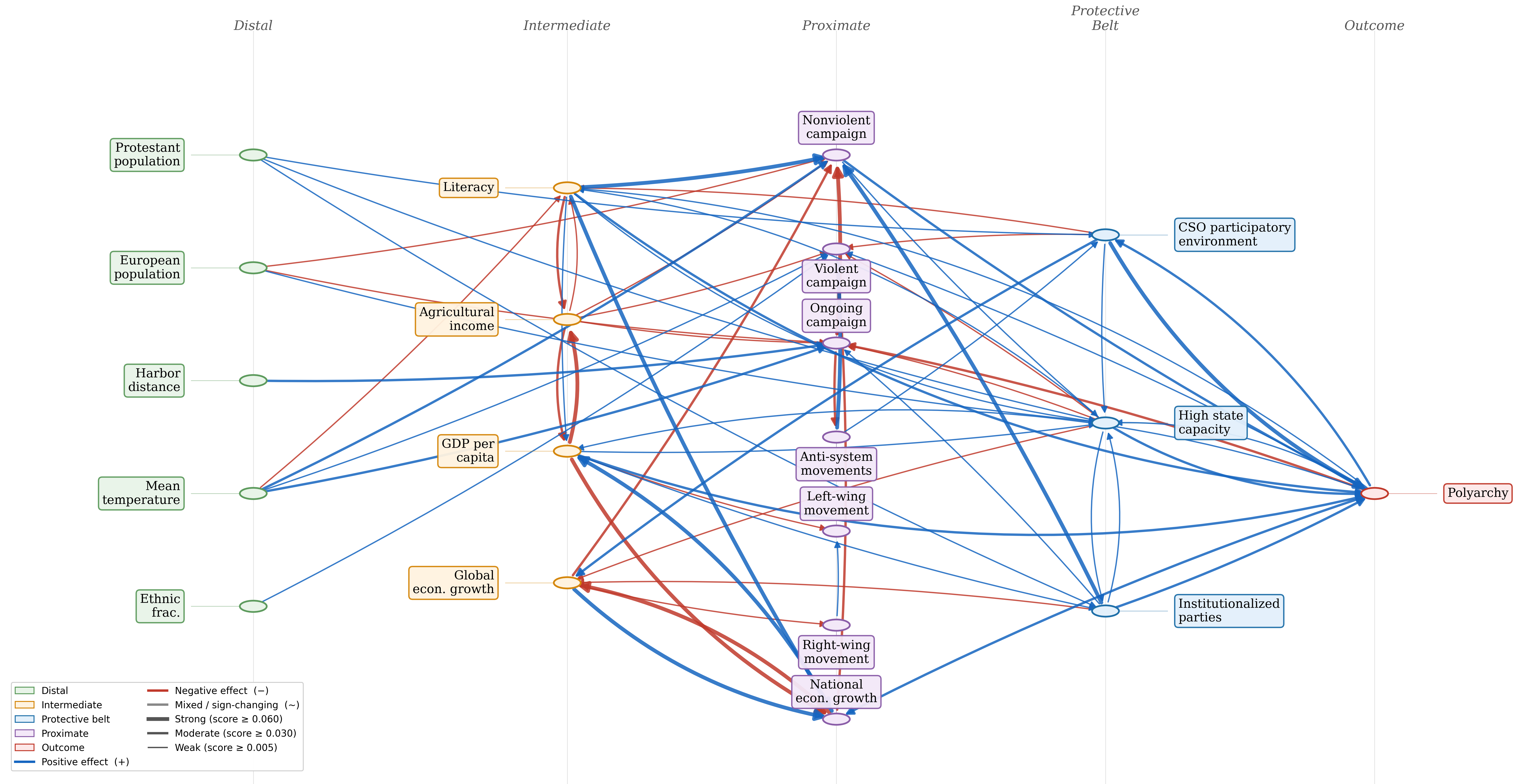}
\caption{The dominant-necessary causal network (84 edges, seed 42).
Nodes are arranged in the causal-sequence layout of the baseline
model's path diagram: structural and distal causes at the left,
then intermediate and proximate/episodic variables, the protective
belt, and the outcome (polyarchy) at the right. Structural variables
appear as sources only; edge width is proportional to the NAVAR
causal score; blue arrows denote positive ICE-derived effects, red
negative, and grey mixed or sign-changing. Edge inclusion is read
directly from the GMM-filtered adjacency matrix
(\texttt{w\_active\_gmm.csv}).}
\label{fig:network}
\end{figure}

Figure~\ref{fig:network} summarizes the estimated graph using the
same conceptual layout as the baseline path diagram, placing
structural and distal conditions on the left, intermediate and
episodic variables in the center, and the protective belt together
with polyarchy on the right. The continuity of the layout makes the
principal similarities and differences immediately visible. The
protective belt remains the dominant set of influences on polyarchy,
confirming the central finding of the baseline model. At the same
time, reciprocal relationships emerge from polyarchy back to the
belt and intermediate institutions, revealing the mutually
reinforcing dynamics that recursive specifications could not
represent. Finally, the outgoing relationships of agricultural
income terminate in the intermediate and episodic layers rather than
at polyarchy itself, visually summarizing the indirect pattern of
influence established in Section~\ref{sec:rq3}.

Table~\ref{tab:comparison} provides a relationship-level comparison
between the baseline model and the proposed workflow. For each
headline relationship, it reports the causal score and direction, the
frequency with which the relationship is retained across the
robustness ensemble, the corresponding coefficient from the baseline
model, the estimated functional form, and the resulting inferential
diagnosis. Viewed together, the figure and table summarize the three
principal conclusions of the analysis: the workflow reproduces the
baseline's core structure, identifies relationships that the linear
model could not represent, and distinguishes among null findings
that appear empirically identical under the linear specification.

\begin{table}[t]
\centering
\caption{Relationship-level comparison: flexible workflow vs.\ linear baseline. Causal scores are unsigned contribution magnitudes from the dominant-necessary graph; signs and functional forms are derived from the functional characterization. Baseline coefficients from
\textcite{coppedge2022}, Table~8.1 and Figure~8.2. The two columns are
not on a common scale; the comparison is between verdicts, not
magnitudes.}
\label{tab:comparison}
\footnotesize
\setlength{\tabcolsep}{4pt}
\begin{tabular}{@{}lcccll@{}}
\toprule
Relationship & \begin{tabular}[c]{@{}c@{}}Causal\\score (sign)\end{tabular}
     & Retention
     & \begin{tabular}[c]{@{}c@{}}Baseline\\coefficient\end{tabular}
     & Functional form & Diagnosis \\
\midrule
\multicolumn{6}{@{}l}{\emph{Protective belt} $\rightarrow$ \emph{polyarchy}} \\
\quad Civil society (CSO)        & 0.073 $(+)$ & 5/5 & 0.0515***    & linear $\to$ sat. & Replicated \\
\quad Rule of law                & 0.039 $(+)$ & 5/5 & 0.0483***    & linear $\to$ sat. & Replicated \\
\quad Party institutionalization & 0.033 $(+)$ & 5/5 & 0.097 (n.s.) & linear $\to$ sat. & Strengthened \\
\addlinespace
\multicolumn{6}{@{}l}{\emph{Modernization variables} $\rightarrow$ \emph{polyarchy}} \\
\quad Literacy rate              & 0.040 $(+)$ & 4/5 & 0.0002 (n.s.) & linear $\to$ sat. & Masked by specification \\
\quad Log GDP per capita         & 0.033 $(+)$ & 4/5 & 0.0137 (n.s.) & sign-changing     & Averaged heterogeneity \\
\quad Agricultural income share  & ---         & 0/5 & $-$0.0014*    & (indirect paths)  & Direct effect dissolves \\
\addlinespace
\multicolumn{6}{@{}l}{\emph{Outside the baseline model}} \\
\quad Nonviolent campaign        & 0.030 $(+)$ & 3/5 & not in model  & hinge             & Suggestive extension \\
\addlinespace
\multicolumn{6}{@{}l}{\emph{Feedback:} \emph{polyarchy} $\rightarrow$ \emph{X (precluded in baseline)}} \\
\quad $\rightarrow$ Civil society (CSO) & 0.047 $(+)$ & 5/5 & not modeled & linear $\to$ sat. & Newly estimable \\
\quad $\rightarrow$ Literacy rate       & 0.023 $(+)$ & 5/5 & not modeled & linear $\to$ sat. & Newly estimable \\
\quad $\rightarrow$ Rule of law         & 0.022 $(+)$ & 5/5 & not modeled & mixed             & Newly estimable \\
\bottomrule
\end{tabular}

\vspace{2pt}
\raggedright\footnotesize
\emph{Note.} ***$p<0.001$, *$p<0.05$; n.s.\ = not significant in the
baseline. Retention = seeds (of five) in which the relationship is
dominant-necessary (\ref{app:ensemble}).
Diagnoses follow the three-way comparison of baseline, benchmark,
and flexible model described in Section~\ref{sec:benchmark-method}.
``Linear $\to$ sat.''\ abbreviates a response that is approximately
linear in low-democracy contexts and saturating at higher democratic
levels; ``hinge'' and ``mixed'' are defined in the text, and
estimated curves for every table relationship appear in~\ref{app:ice}.
\end{table}

\section{Scope and Limitations}
\label{sec:scope}

The workflow is designed for panel time series in which dynamic
relationships may be nonlinear, reciprocal causation is
theoretically plausible, structural conditions coexist with evolving
institutions, and the research question concerns not only whether
one variable influences another but how that influence changes
across political contexts. These features characterize many
substantive problems in comparative politics, including
democratization, institutional change, conflict, and economic
development. Nothing in the workflow, however, depends specifically
on V-Dem or on the causal sequence model examined here.

The workflow is intended to strengthen empirical inference, not to
replace established principles of causal identification. Forecast
necessity should therefore not be interpreted as evidence of
counterfactual causation. The design incorporates no exogenous
shocks, no instrumental variables, and no identifying assumptions
beyond those of the autoregressive model class itself. Consequently,
relationships that are necessary for accurate prediction may still
reflect unobserved confounding. The estimated graph is therefore best
understood as a disciplined representation of a system's dynamic
dependencies and a demanding empirical benchmark against which
proposed causal explanations can be evaluated, rather than as a
substitute for design-based causal inference.

Neural estimation also introduces Monte Carlo variation, making
estimation stability itself an object of inference. All reported
results derive from a documented reference seed and are accompanied
by a five-seed robustness ensemble (\ref{app:ensemble}). The principal findings are highly stable, whereas
relationships with lower ensemble agreement are reported as
provisional rather than definitive. A separate limitation concerns
resolution rather than stability. Functional forms are characterized
within terciles of lagged polyarchy, so conclusions regarding regime
dependence inherit the coarseness of that partition. Richer
representations of state dependence are possible in principle but
lie beyond the scope of the present analysis.

A related measurement caveat concerns the scale of the indices
themselves. V-Dem's high-level indices are bounded and constructed
through a transformation that compresses variation near the ceiling
\parencite{pemstein2018}, and compression of the target in
high-democracy contexts could contribute to the saturation estimated
there. Two observations bound the concern without eliminating it.
Standardization does not address it, since linear rescaling leaves
the compression intact; and a purely mechanical ceiling effect would
flatten every relationship into the same bounded target alike,
whereas the estimated forms include sign-changing, hinge, and mixed
responses alongside the saturating ones. Re-estimating the
characterization on an unbounded transformation of the indices is a
natural robustness exercise, and we flag it as such rather than
claim the question settled. The estimated graph also admits a time-varying extension, with impulse-response and counterfactual machinery, developed in
\textcite{kdd2026dcnar}.

These boundaries are central rather than incidental. The
workflow's contribution is not to infer causation from prediction
but to improve the empirical evaluation of dynamic theories:
political theories can be tested against a model class capable of
representing the mechanisms they propose, with relationship-level
interpretation and pre-committed standards of evidence preserved.

\section{Conclusion}
\label{sec:conclusion}

This article began with a simple inferential question: what does a
null relationship in a dynamic panel model actually mean? Within the
standard linear toolkit, the answer is fundamentally ambiguous. A
non-significant coefficient may reflect a genuinely absent
relationship, heterogeneous effects averaged toward zero, reciprocal
dynamics that the model cannot represent, or relationships obscured
by a reduced specification. We have presented a workflow that
distinguishes among these possibilities by combining flexible
autoregressive estimation, forecast-necessity testing, functional
characterization, and same-data linear benchmarking within a single
inferential framework. Applied to the causal sequence model of
democratization, the workflow demonstrates that these different
sources of apparent null findings can be identified empirically
rather than inferred by interpretation alone.

The protective belt of civil society, the rule of law, and
institutionalized parties is recovered under a substantially more
flexible model class, and its stabilizing influence proves
strongest where democracy is least consolidated. Reciprocal relationships running from democracy to its
institutional foundations, long central to theories of democratic
consolidation but excluded by recursive specifications, become both
estimable and necessary for accurate prediction. Most importantly, three weak direct effects that give no sign, in
the baseline model, of being weak for different reasons receive
three distinct diagnoses: one relationship removed from the
published record, and two recovered, demonstrating that the workflow's
corrections run in both directions rather than uniformly toward
finding more.
More broadly, the article argues that the evaluation of dynamic
political theories should be limited as little as possible by the
assumptions of the estimation framework. Many influential theories of
political development posit nonlinear relationships, reciprocal
dynamics, and complex interactions among highly correlated
institutions, whereas the workhorse models used to evaluate them
often exclude precisely those features. The workflow presented here
narrows that gap while preserving the qualities that have made the
linear toolkit valuable: relationship-level interpretability,
transparent empirical testing, and explicit inferential standards.
The principal contribution of the workflow is therefore not merely
different estimates but a determination of what each estimate
represents, and, where the published record cannot be sustained
under the full system, a disciplined account of why.

\paragraph{Funding statement.} 
This work was supported in part by funding from the Kellogg
Institute for International Studies, the Franco Family Institute for
Liberal Arts and the Public Good, and the Lucy Family Institute
for Data \& Society at the University of Notre Dame.
\paragraph{Data availability statement.} Replication materials,
including the input panel, all estimation and analysis code, the
trained model checkpoint underlying every reported number, reference
outputs for each stage, and code reproducing all figures, will be
deposited in the Political Analysis Dataverse upon conditional
acceptance, in accordance with the journal's replication policy. The
V-Dem source data are publicly available from the V-Dem Institute.
\paragraph{Competing interests.} The authors declare none.
\paragraph{Author contributions.}  Conceptualization: M.C., D.Z.; Methodology: D.Z., V.K.; Software and formal analysis: V.K.,
D.Z.; Data curation: V.K.; Writing, original draft: V.K., D.Z.,
M.C.; Writing, review and editing: M.C., D.Z., V.K. All authors
approved the final submitted draft.
\paragraph{AI use.} In this work, generative AI tools were used for coding assistance and to polish author-written text. Authors take full responsibility for all contents of the article.

\printbibliography

\appendix

\section{Variables}
\label{app:variables}
Table~\ref{tab:variables} lists the 23 variables of the estimated
system in model order, with the V-Dem or source dataset from which
each is drawn and its conceptual category in the baseline model's
framework. Four of the variables were dropped from the original dataset of 27 because of no within-country variation (non-middle-sector campaign, non-worker-farmer campaign, worker-farmer campaign, middle-sector campaign). The five variables categorized as structural are those
designated exogenous by the ICC screen of Section~3.1 of the main
text; they enter the analysis as candidate causes only.
\begin{table}[!htbp]
\centering
\caption{The 23 variables of the final model. Categories follow
\textcite{coppedge2022}.}
\label{tab:variables}
\small
\begin{tabular}{@{}llll@{}}
\toprule
Variable & Code & Source & Category \\
\midrule
Electoral democracy (polyarchy) & v2x\_polyarchy & V-Dem v9 & Outcome \\
CSO participatory environment & v2csprtcpt & V-Dem v9 & Protective belt \\
Rule of law / state capacity & v2clrspct & V-Dem v9 & Protective belt \\
Institutionalized parties & v2xps\_party & V-Dem v9 & Protective belt \\
Majoritarian electoral system & elecsys & V-Dem v9 & Intermediate \\
Anti-system movement & v2csantimv & V-Dem v9 & Proximate \\
Left-wing movement & v2csanmvch\_6 & V-Dem v9 & Proximate \\
Right-wing movement & v2csanmvch\_7 & V-Dem v9 & Proximate \\
National economic growth & ecgrowth & V-Dem derived & Proximate \\
Violent campaign & navco\_viol & NAVCO 1.1 & Proximate \\
Nonviolent campaign & navco\_nonviol & NAVCO 1.1 & Proximate \\
Global economic growth & expand & V-Dem derived & Intermediate \\
Strict presidentialism & hardpres & \textcite{coppedge2022}, Ch.~6 & Intermediate \\
Ongoing campaign & dkw\_ongoing & DKW & Proximate \\
Log GDP per capita & e\_migdppclni & V-Dem derived & Intermediate \\
Literacy rate & litcyxtend & \textcite{coppedge2022}, Ch.~4 & Intermediate \\
Agricultural income share & s\_mil\_agro & \textcite{miller2015} & Intermediate \\
Resource dependence & s\_mil\_resdep2 & \textcite{miller2015} & Proximate \\
Protestant population & chrstprotpct & \textcite{coppedge2022}, Ch.~3 & Structural \\
European-descended population & eur\_pct\_est\_smooth & \textcite{coppedge2022}, Ch.~3 & Structural \\
Natural harbor distance & portdist\_natural\_km & \textcite{coppedge2022}, Ch.~3 & Structural \\
Mean temperature & tmean & \textcite{coppedge2022}, Ch.~3 & Structural \\
Ethnic fractionalization & wim\_ethfrac & \textcite{coppedge2022}, Ch.~3 & Structural \\
\bottomrule
\end{tabular}

\vspace{2pt}
\raggedright\footnotesize
\emph{Note.} NAVCO 1.1 = \textcite{chenoweth2011}; DKW =
\textcite{dahlum2019}. Structural variables are ICC-masked as targets
(Section~3.1 of the main text); ICC values in
Section~4 of the main text.
\end{table}

The baseline specification \parencite{coppedge2022} retained 10 of 27 candidate variables. Multicollinearity constrained how many correlated institutional
measures could be evaluated by the model jointly, while other candidates, such as economic
growth and social movement campaigns among them, were set aside on
theoretical grounds as predictors of upturns and downturns rather
than of levels. Reciprocal effects were entered into the baseline only as
lagged cross-paths within a recursive ordering. Many were tested,
few reached significance, and only those few appear in the published
study, inlcuding a negative lagged effect of polyarchy on
institutionalized parties. In the present re-estimation, the
excluded variables re-enter as potential causes, potential outcomes,
and potential mediators of the dynamic relationships under study.

\emph{Panel construction.} Countries with fewer than 9 observed years
are excluded (one lag window requires 9 years at $K=8$). The final
five years of each country's series are reserved at panel construction
and excluded from all training.

The raw extract comprises 18{,}604 country-year rows. After rows with incomplete covariate data are dropped and countries with fewer than nine usable years are
excluded, the panel comprises 5{,}166 observations on 113 countries,
1900--2001, with all variables standardized. Up to five final years
of each country's series, 485 observations in all, are withheld from
all estimation, leaving 4{,}681 country-year observations for the
analysis.

\section{Existing Remedies and Extended Justifications}
\label{app:remedies}

This appendix develops three points that the main text states in
compressed form: why the standard remedies for nonlinearity within
linear models do not resolve the inferential ambiguity, why the
standard remedies for dimensionality do not either, and why the
dominance threshold defers to the empirical distribution of loss
differentials.

\subsection{Nonlinearity Within Linear Models}

Linearity in parameters does not, of course, preclude nonlinearity in variables. Analysts routinely represent conditional and curved relationships within linear models through multiplicative interactions, polynomial terms, and transformed variables, and a substantial methodological literature governs their use \parencite{brambor2006}. The difficulty is that every such remedy requires the analyst to specify, before estimation, which relationships bend, on which moderating variable, and in what functional form. Even in the simplest case of a single interaction, standard practice is fragile: multiplicative models impose a linear-interaction assumption that frequently fails in published applications, and estimates can be badly misleading when the true conditioning is nonlinear or local \parencite{hainmueller2019}. In a dynamic system of twenty or more jointly evolving variables, the space of candidate interactions and transformations grows combinatorially, the appropriate moderator is itself a research question, and pre-specifying the functional form amounts to assuming what the analysis is meant to discover. The practical consequence is that nonlinearity, though representable in principle, is routinely omitted or misspecified in dynamic applications. This observation has motivated calls for more flexible functional forms in political analysis since at least \textcite{beck2000}.

\subsection{Aggregation and Regularization}

Omission is not the only remedy. Correlated predictors are sometimes replaced by a smaller number of estimated components, whether extracted factor scores or reflective latent variables in a structural equation framework \parencite{bollen1989, jackman2008}, and dynamic factor models extend the same logic to panel time series \parencite{stock2016}. These approaches stabilize estimation, and where the correlated indicators genuinely measure a single underlying construct, they are the right tool. Their cost appears when the indicators are theoretically distinct: a composite absorbs its components, and relationship-level questions about the components can no longer be asked. A model in which an institutional factor predicts democratization cannot say whether civil society, the rule of law, or party institutionalization carries the effect, in which direction each operates, or whether democracy feeds back on one but not another. For a theory whose claims are stated at the level of distinct institutions, as the causal sequence model's are, aggregation purchases stability by giving up precisely the quantities under study. Regularized estimators preserve the individual variables and shrink their coefficients instead, but shrinkage stabilizes estimation without indicating which retained relationships are necessary rather than merely tolerated by the penalty.

\section{Structural exogeneity and asymmetric masking} 
\label{app:strexo}
Geography, colonial history, religious
composition, and other deep structural conditions shape political
development but are not themselves products of the political dynamics
under study. A fully symmetric model therefore allocates capacity to
relationships that cannot plausibly exist; for example, allowing
contemporary political institutions to predict a country's
temperature or distance from the sea. Simply removing these
variables, however, discards information that is often essential for
explaining political outcomes. The workflow resolves this tension by
distinguishing the two roles a variable can play: structural
covariates enter the model as potential causes but are excluded as
candidate outcomes.

\begin{figure}[!htbp]
\centering
\includegraphics[width=0.9\textwidth]{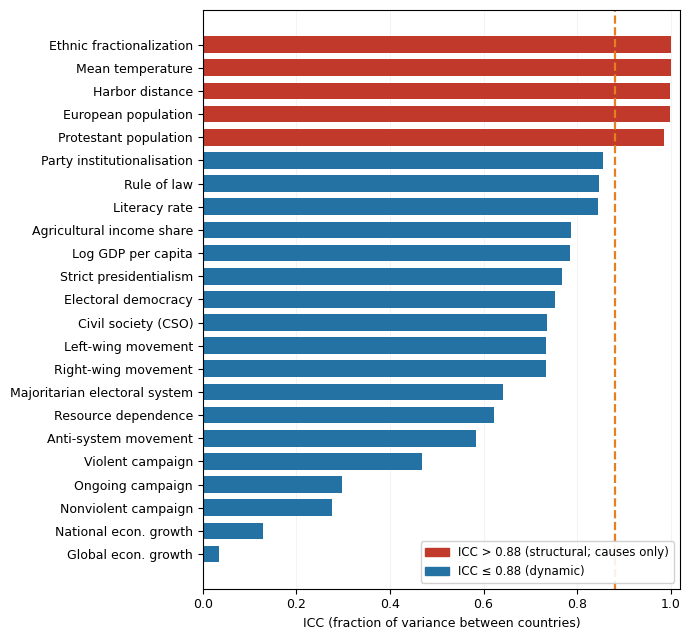}
\caption{The structural exogeneity screen. Bars show the intraclass
correlation coefficient of each model variable, the fraction of its
variance lying between countries rather than within them. The five
variables above the 0.88 threshold (red) have essentially no
within-country movement over the observation period and are
designated structurally exogenous: they enter the analysis as
candidate causes but are never modeled as outcomes. The threshold
falls within a wide empty interval, between 0.856 (party
institutionalization) and 0.987 (Protestant population share), so
the classification does not depend on its exact placement.
Slow-moving institutional stocks (party institutionalization, rule
of law, literacy) sit high among the dynamic variables; episodic
variables (campaigns, growth) sit lowest.}
\label{fig:icc}
\end{figure}

The distinction is operationalized through the data rather than
through substantive judgment. Variables whose variation lies almost
entirely between countries, with essentially no within-country
movement over the period of observation, cannot plausibly be treated
as outcomes of annual political dynamics, regardless of their role as
causes. We therefore classify variables using their intraclass (within-country)
correlation coefficient (ICC) and designate those above a high
threshold as structurally exogenous ($\mathrm{ICC}>0.88$ in the
present application, where the criterion cleanly separates the variable
set). Once designated, these variables remain fully active as causes:
their influence on every dynamic outcome is estimated through a
contribution function, tested for forecast necessity, and
characterized in functional form, exactly as for any other candidate
relationship (Figure~\ref{fig:icc}). What the designation suppresses is the reverse
direction. No equation is estimated for a structural variable
itself, so no contribution function points into one, no
forecast-necessity test concerns its prediction, and no response
curve is constructed with a structural variable as the outcome. The
treatment is the nonlinear analogue of including exogenous
regressors in a VARX model, which enter the right-hand side of every
equation but receive no equation of their own
\parencite{lutkepohl2005}.

The asymmetry is imposed during estimation rather than by discarding estimated relationships afterward, ensuring that the model's capacity is devoted entirely to relationships that are theoretically admissible. Apart from the initial empirical screen, the procedure requires no analyst discretion and generalizes directly to panel applications that combine slowly varying structural conditions with rapidly evolving institutional variables. Figure~\ref{fig:icc} displays the full ICC distribution: the five
designated variables are separated from the dynamic set by a wide
empty interval, so the screen involves no borderline judgment.

\section{Estimation Details}
\label{app:estimation}
\emph{NAVAR.} Lags $K=8$; one hidden layer of 32 nodes per
contribution network; dropout 0.10; $\ell_1$ penalty $\lambda_1 =
0.15$; weight decay $10^{-3}$; batch size 128; Adam with learning rate
$3 \times 10^{-4}$; variables standardized before training. Training
is 1{,}000 epochs total: a 250-epoch warm start with the standard
NAVAR objective, followed by 750 epochs of masked fine-tuning at half
the learning rate, in which structural target columns are excluded
from both the prediction loss and the $\ell_1$ term. Model selection
is by checkpoint on masked validation loss over a terminal 10\%
window split.

\emph{Sign determination.} The same curves determine each relationship's direction. Within each regime, the estimated response has a signed slope; the net effect of a relationship is the aggregate of these slopes across regimes, and its sign is the direction that prevails. Because the slopes are computed regime by regime, the procedure also yields a measure of sign stability: a relationship whose direction agrees across all three democratic contexts is directionally unambiguous even when its strength varies. In the application below, every retained relationship meets this standard, so the signs reported in Section~5 of the main text are properties of the system rather than artifacts of any single regime.

\emph{Seed.} All main-text values use seed 42;~\ref{app:ensemble} reports the five-seed ensemble. The
replication package contains the full configuration object, the
ensemble driver, and per-seed outputs.

\section{Forecast- and Dominance-Necessity Construction}
\label{app:fnc}
\subsection{Forecast Necessity}
The estimation panel of
4{,}681 country-year observations yields 3{,}777 prediction windows,
each pairing eight years of lagged values with the observation they
precede; the final 755, twenty percent of the total, form a fixed
evaluation set used for every test. No re-estimation is involved:
the model is fitted once, and each of the 396 candidate
relationships is ablated exactly, by subtracting its contribution
function from the fitted predictions, before full and restricted
forecasts are compared on the evaluation windows. The comparison is
quantified by the loss differential
\begin{equation}
d_{ij,t} \;=\;
\underbrace{L^{(-ij)}_{t}}_{\substack{\text{forecast loss}\\ \text{without the relationship}}}
\;-\;
\underbrace{L_{t}}_{\substack{\text{forecast loss}\\ \text{of the full model}}},
\label{eq:dm}
\end{equation}
which is positive when removing the relationship makes forecasts
worse. Loss differentials are pooled across countries, and the
Diebold--Mariano statistic follows the standard construction for
one-step-ahead forecasts (HAC lag 0); the test is one-sided at
$\alpha = 0.05$. The criterion is directional and target-specific:
a variable may be necessary for predicting one outcome and redundant
for another. Temporal ordering is preserved throughout the
evaluation. The evaluation windows are the last segment of the
pooled panel rather than the terminal years of each country's
series, and a portion of them lies within the span used to fit the
model; because the full and restricted models share identical
weights and are compared on identical windows, the comparison
isolates the contribution of the ablated relationship.
~\ref{app:ensemble} repeats the analysis on a strictly
post-training evaluation set, and every principal finding persists,
with attrition confined to relationships reported as provisional. A
further five years per country were removed before any estimation
and are used nowhere in the analysis.

\subsection{Dominance-Necessity}
Deferring to the empirical distribution here is a considered choice, not a retreat from theory. Theoretical guidance operates at the level of which variables belong in the system and which relationships are admissible: commitments the workflow honors in its variable set, its exogeneity screen, and its additive structure, but no theory of democratization speaks to the magnitude of forecast-loss differential that separates a marginal relationship from an organizing one. For a decision at that resolution, the alternatives are a threshold chosen by the researcher, which would be neither theoretically grounded nor free of discretion, and a threshold learned from the structure of the loss differentials themselves. Where theory is silent, the distribution is the more defensible guide. The procedure exercises discretion once, in fixing the form of
the rule (a two-component mixture on the logarithm of the loss
differentials), and never in placing the line; the bimodal structure
the rule presumes appears under every seed of the robustness
ensemble, as described below.

\emph{Dominance filter} is a two-component Gaussian mixture
($\texttt{n\_init}=10$) on $\log_{10}$ mean loss differentials among
necessary edges. Threshold is at the rightmost component intersection (Figure~\ref{fig:gmm}).

\section{Edge Testing and the Full Edge Table}
\label{app:edges}

\begin{figure}[!htbp]
\centering
\includegraphics[width=\textwidth]{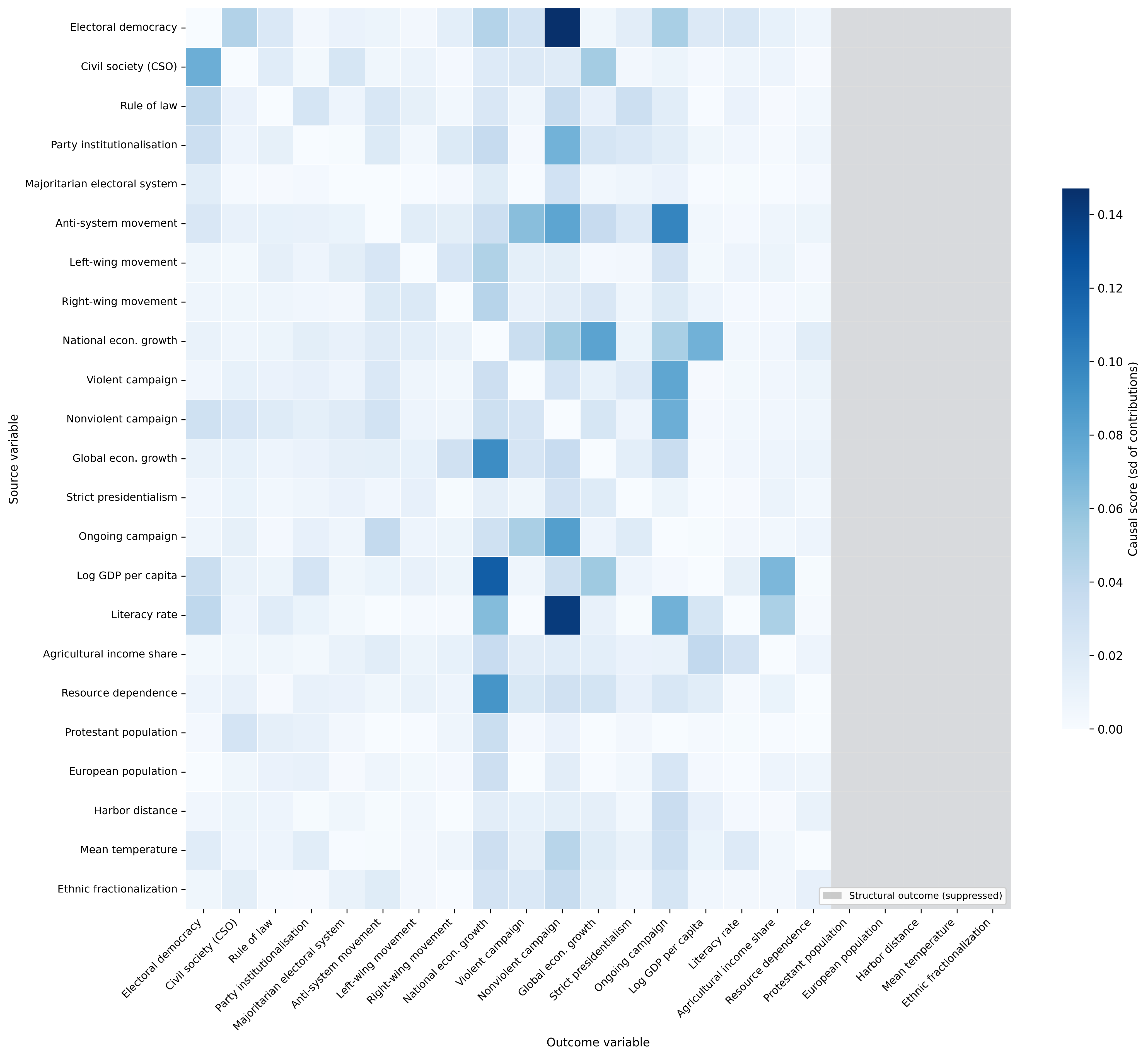}
\caption{Estimated causal score matrix. Each cell reports how
actively a source variable (rows) moves the prediction of an outcome
variable (columns), before necessity testing. Shaded columns mark
the five structurally exogenous variables, which enter as candidate
causes only. The concentration of scores in a small share of cells
reflects the sparsity that the $\ell_1$ penalty induces; the columns
for polyarchy and its institutional supports carry the strongest
signal. The complete matrix is reported numerically in
Table~\ref{tab:scorematrix}.}
\label{fig:scorematrix}
\end{figure}

Figure~\ref{fig:scorematrix} displays the estimated causal score
matrix from which the analysis begins; the remainder of this
appendix traces the path from these candidate scores to the
estimated graph. Table~\ref{tab:scorematrix} reports the complete matrix with both necessity tiers marked, so the full path from candidate scores to the estimated graph is inspectable relationship by relationship. Two quantities supplement that figure
for exact reproduction. Retained mean loss differentials span
$5.95 \times 10^{-4}$ to $3.67 \times 10^{-2}$, and the largest
excluded differential is $5.12 \times 10^{-4}$, so verification of
a re-estimated threshold is a matter of range membership rather
than point agreement.  Figure~\ref{fig:gmm} displays the criterion itself: the fitted
mixture, the placement of the threshold, and the ranked loss
differentials it partitions.

\begin{figure}[htbp]
\centering
\includegraphics[width=\textwidth]{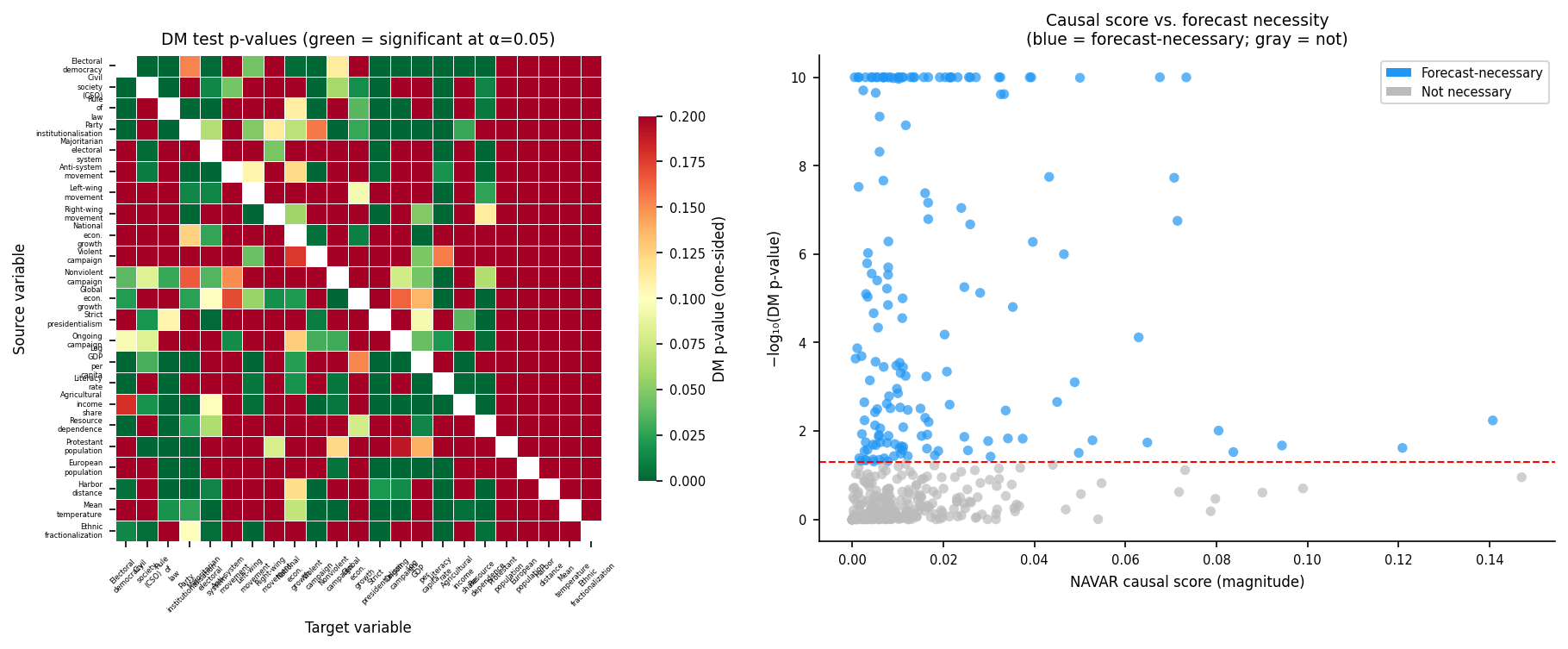}
\caption{Forecast-necessity testing across the 396 candidate
relationships. Left: one-sided Diebold--Mariano $p$-values for the
exact ablation of each source (rows) from each target equation
(columns); green cells are significant at $\alpha = 0.05$, blank
cells are excluded self-effects and masked structural targets, and
$p$-values above 0.2 share the terminal color. Right: each
candidate's $-\log_{10}$ DM $p$-value against its causal score
magnitude, with the dashed line at $\alpha = 0.05$ and $p$-values
floored at $10^{-10}$ for display; blue points are the 160
forecast-necessary relationships. Necessity is not a monotone
function of score: candidates with some of the largest scores fall
below the line, and candidates with small scores clear it.}
\label{fig:dmtests}
\end{figure}

\begin{figure}[!htbp]
\centering
\includegraphics[width=\textwidth]{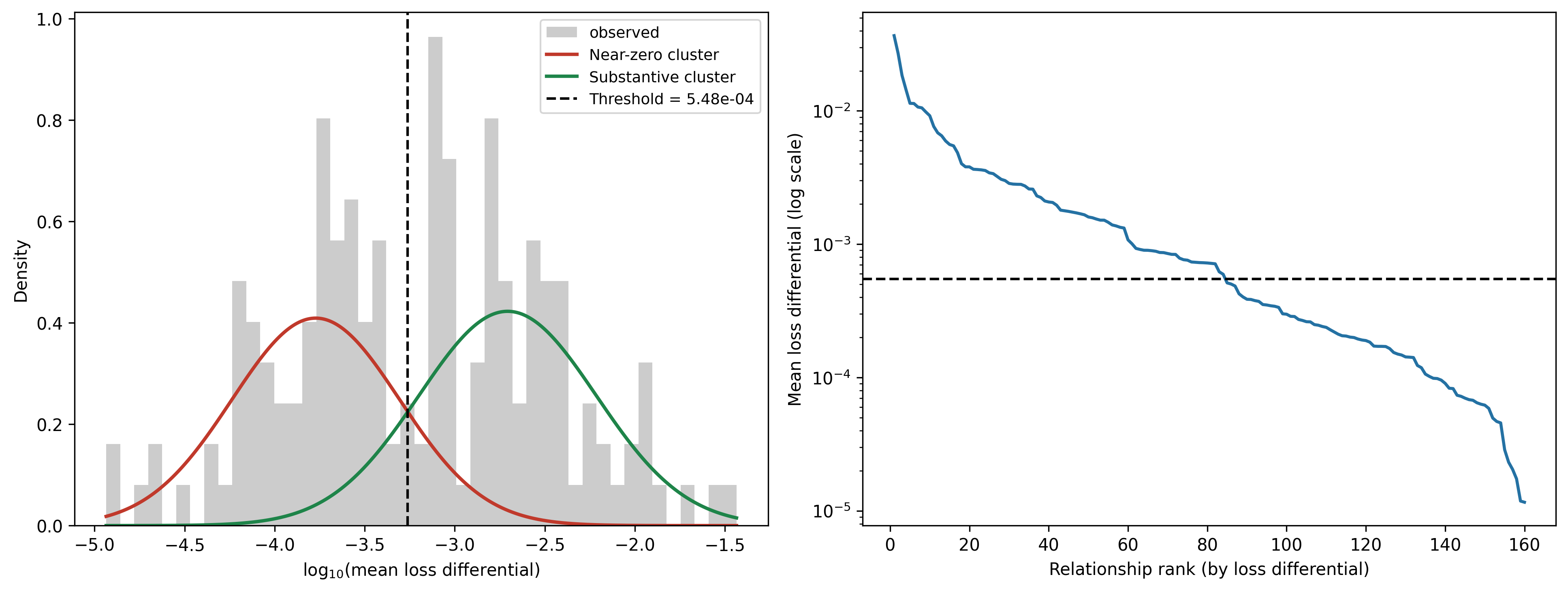}
\caption{The automatic dominance criterion. Left: a two-component
mixture fitted to the logarithm of mean loss differentials across
the 160 forecast-necessary relationships separates a near-zero
cluster from a substantive one; the dashed line marks their
intersection, $5.48 \times 10^{-4}$, which serves as the retention
threshold. Right: the same 160 relationships ranked by loss
differential, with 84 above the threshold. No researcher-specified
parameter enters the decision.}
\label{fig:gmm}
\end{figure}

Figure~\ref{fig:dmtests} reports the necessity testing itself. The
left panel maps the one-sided DM $p$-value of every candidate
ablation; the right panel plots each candidate's $p$-value against
its causal score magnitude. The two quantities are related but far
from interchangeable: the largest causal score in the system, the
0.147 score of polyarchy on nonviolent campaigns, fails the
necessity test, while relationships with scores an order of
magnitude smaller pass it. The pattern is the empirical counterpart
of the motivation in Section~3.3 of the main text: with correlated
predictors, a causal score measures activity, not indispensability,
and retention therefore requires the ablation test.

The remaining per-relationship
quantities, including DM $p$-values, mean loss differentials, and the sign
and stability of each characterized response, are provided in the
replication archive (see the Data Availability statement of the main
text), which also contains the 84-relationship adjacency matrix
used for all main-text figures.

\begin{sidewaystable}
\centering
\caption{Estimated causal score matrix (complete), with
necessity tiers. Entries are causal scores
$S_{ij} \times 100$; rows are sources, columns dynamic
outcomes. Bold entries mark the 84 dominant-necessary
relationships constituting the estimated graph; daggered
entries mark the further 76 relationships that are
forecast-necessary but fall below the dominance threshold;
plain entries are not necessary for out-of-sample
prediction. The five structurally exogenous variables (final
five rows) enter as sources only; self-effects are excluded
(dashes). This table reports the matrix shown as Figure~2 of
the main text; variable codes are defined in
Appendix~\ref{app:variables}.}
\label{tab:scorematrix}
\scriptsize
\setlength{\tabcolsep}{2.5pt}
\begin{tabular}{@{}lrrrrrrrrrrrrrrrrrr@{}}
\toprule
Source & \rotatebox{60}{\texttt{v2x\_polyarchy}} & \rotatebox{60}{\texttt{v2csprtcpt}} & \rotatebox{60}{\texttt{v2clrspct}} & \rotatebox{60}{\texttt{v2xps\_party}} & \rotatebox{60}{\texttt{elecsys}} & \rotatebox{60}{\texttt{v2csantimv}} & \rotatebox{60}{\texttt{v2csanmvch\_6}} & \rotatebox{60}{\texttt{v2csanmvch\_7}} & \rotatebox{60}{\texttt{ecgrowth}} & \rotatebox{60}{\texttt{navco\_viol}} & \rotatebox{60}{\texttt{navco\_nonviol}} & \rotatebox{60}{\texttt{expand}} & \rotatebox{60}{\texttt{hardpres}} & \rotatebox{60}{\texttt{dkw\_ongoing}} & \rotatebox{60}{\texttt{e\_migdppclni}} & \rotatebox{60}{\texttt{litcyxtend}} & \rotatebox{60}{\texttt{s\_mil\_agro}} & \rotatebox{60}{\texttt{s\_mil\_resdep2}} \\
\midrule
\texttt{v2x\_polyarchy} & -- & \textbf{4.6} & \textbf{2.2} & 0.4 & \textbf{1.0} & 0.8 & 0.5$^{\dagger}$ & 1.5 & \textbf{4.5} & \textbf{2.8} & 14.7 & 0.6 & \textbf{1.6} & \textbf{5.0} & \textbf{2.0} & \textbf{2.3} & 1.2$^{\dagger}$ & 0.7$^{\dagger}$ \\
\texttt{v2csprtcpt} & \textbf{7.3} & -- & \textbf{1.7} & 0.4 & \textbf{2.5} & 0.6$^{\dagger}$ & 0.9 & 0.3 & 1.9 & \textbf{2.0} & 1.8 & \textbf{5.3} & \textbf{0.4} & 0.9 & 0.3 & \textbf{0.7} & 0.8 & 0.2$^{\dagger}$ \\
\texttt{v2clrspct} & \textbf{3.9} & 1.0 & -- & \textbf{2.6} & \textbf{0.8} & 2.3 & 1.3 & 0.5 & 2.3 & \textbf{0.7} & 3.6 & \textbf{1.2} & \textbf{3.3} & \textbf{1.6} & 0.1 & \textbf{1.0} & 0.2 & 0.5$^{\dagger}$ \\
\texttt{v2xps\_party} & \textbf{3.3} & 0.8 & \textbf{1.3} & -- & 0.1 & 2.0 & 0.5$^{\dagger}$ & 2.0 & 3.7 & 0.3 & \textbf{7.1} & \textbf{2.5} & \textbf{2.2} & \textbf{1.7} & 0.6$^{\dagger}$ & 0.6$^{\dagger}$ & 0.3$^{\dagger}$ & 0.6 \\
\texttt{elecsys} & 1.6 & 0.3$^{\dagger}$ & 0.2 & 0.3 & -- & 0.0 & 0.1 & 0.3$^{\dagger}$ & 1.8 & 0.1 & 2.8 & 0.5 & \textbf{0.7} & 1.0 & 0.1 & 0.1$^{\dagger}$ & 0.1 & 0.3$^{\dagger}$ \\
\texttt{v2csantimv} & 2.3 & \textbf{1.1} & 1.2 & 1.1$^{\dagger}$ & \textbf{1.0} & -- & 1.6 & 1.5 & 3.2 & \textbf{6.3} & 8.0 & 3.7 & \textbf{2.1} & 9.9 & 0.5 & 0.3$^{\dagger}$ & 0.6 & 0.8$^{\dagger}$ \\
\texttt{v2csanmvch\_6} & 0.6 & 0.4 & 1.4 & 0.8$^{\dagger}$ & \textbf{1.5} & 2.4 & -- & 2.4 & 4.7 & 1.4 & 1.5 & 0.3 & 0.4 & 2.7 & 0.4 & 0.8$^{\dagger}$ & 0.8 & 0.3$^{\dagger}$ \\
\texttt{v2csanmvch\_7} & 0.7 & 0.6 & 0.7 & 0.5$^{\dagger}$ & 0.5 & 2.0 & \textbf{2.1} & -- & 4.4 & 1.1 & 1.6 & 2.3 & \textbf{0.7} & 2.0 & 0.8$^{\dagger}$ & 0.3$^{\dagger}$ & 0.3 & 0.4 \\
\texttt{ecgrowth} & 1.1 & 0.7 & 0.8 & 1.5 & \textbf{1.1} & 1.8 & 1.4 & 1.0 & -- & \textbf{3.4} & 5.4 & \textbf{8.0} & 0.9 & 5.0 & \textbf{7.1} & 0.5 & 0.5 & 1.6 \\
\texttt{navco\_viol} & 0.5 & 1.2 & 1.0 & 1.2 & 0.8 & 2.1 & 0.7$^{\dagger}$ & 0.5 & 3.2 & -- & 2.6 & 1.2 & 2.0 & 7.9 & 0.2$^{\dagger}$ & 0.4 & 0.5 & 0.8 \\
\texttt{navco\_nonviol} & \textbf{3.0} & 2.4 & \textbf{1.9} & 1.4 & \textbf{1.8} & 2.8 & 0.8 & 0.7 & 3.2 & 2.5 & -- & 2.5 & 0.8 & 7.3 & 0.3$^{\dagger}$ & 0.5$^{\dagger}$ & 0.5 & 0.7 \\
\texttt{expand} & 1.1$^{\dagger}$ & 1.2 & 0.8 & 1.0$^{\dagger}$ & 1.3 & 1.4 & 1.2 & \textbf{3.0} & \textbf{9.4} & 2.5 & \textbf{3.5} & -- & 1.5 & 3.5 & 0.3 & 0.6$^{\dagger}$ & 0.8 & 0.9$^{\dagger}$ \\
\texttt{hardpres} & 0.5 & 0.9$^{\dagger}$ & 0.5 & 0.7 & 1.0$^{\dagger}$ & 0.5 & 1.2 & 0.2 & 1.4 & 0.6$^{\dagger}$ & 2.7 & 1.9 & -- & 0.8 & 0.1 & 0.2 & 0.9$^{\dagger}$ & 0.5$^{\dagger}$ \\
\texttt{dkw\_ongoing} & 0.6 & 1.3 & 0.3 & 1.2 & 0.7 & \textbf{3.7} & 0.8 & 0.8 & 3.0 & \textbf{5.0} & \textbf{8.4} & 0.8 & 1.8 & -- & 0.2$^{\dagger}$ & 0.5$^{\dagger}$ & 0.5 & 0.8$^{\dagger}$ \\
\texttt{e\_migdppclni} & \textbf{3.3} & 1.1$^{\dagger}$ & \textbf{0.8} & \textbf{2.6} & 0.6 & 0.9 & \textbf{1.1} & 0.9 & \textbf{12.1} & 0.7 & 3.2 & 5.5 & 0.8$^{\dagger}$ & 0.3$^{\dagger}$ & -- & 1.3 & \textbf{6.8} & 0.1 \\
\texttt{litcyxtend} & \textbf{4.0} & 0.8 & \textbf{1.7} & 1.0 & 0.4 & 0.0 & 0.3$^{\dagger}$ & 0.2 & \textbf{6.5} & 0.1 & \textbf{14.1} & 1.1 & 0.1$^{\dagger}$ & 7.2 & \textbf{2.5} & -- & \textbf{4.9} & 0.2$^{\dagger}$ \\
\texttt{s\_mil\_agro} & 0.4 & 0.6$^{\dagger}$ & 0.6$^{\dagger}$ & 0.4$^{\dagger}$ & 1.1 & 1.7 & 0.8$^{\dagger}$ & 1.2 & 3.6 & \textbf{1.6} & \textbf{1.7} & 1.5 & \textbf{1.0} & \textbf{1.0} & \textbf{3.9} & \textbf{2.7} & -- & 0.8$^{\dagger}$ \\
\texttt{s\_mil\_resdep2} & 0.8$^{\dagger}$ & 1.1 & 0.2$^{\dagger}$ & 1.1$^{\dagger}$ & 1.0 & 0.6 & 1.1 & 0.8 & 9.0 & 2.2 & 3.0 & 2.7 & 1.2 & 2.3 & 1.6$^{\dagger}$ & 0.3 & 0.9 & -- \\
\midrule
\texttt{chrstprotpct} & 0.3 & \textbf{2.6} & \textbf{1.3} & \textbf{1.1} & 0.4 & 0.0 & 0.1 & 0.7 & 3.4 & 0.3 & 1.0 & 0.0 & 0.4 & 0.0 & 0.3 & 0.1 & 0.1 & 0.0 \\
\texttt{eur\_pct\_est\_smooth} & 0.0 & 0.6 & \textbf{1.0} & 1.1$^{\dagger}$ & 0.2 & 0.7 & 0.4 & 0.3 & 3.2 & 0.1 & \textbf{1.6} & 0.1 & 0.5$^{\dagger}$ & \textbf{2.4} & 0.3$^{\dagger}$ & 0.1$^{\dagger}$ & 0.8 & 0.6 \\
\texttt{portdist\_natural\_km} & 0.5$^{\dagger}$ & 0.9 & 0.8$^{\dagger}$ & 0.1$^{\dagger}$ & 0.6$^{\dagger}$ & 0.1 & 0.5 & 0.1 & 1.6 & 1.2$^{\dagger}$ & 1.4 & 1.4 & 0.5$^{\dagger}$ & \textbf{3.4} & 1.2 & 0.3$^{\dagger}$ & 0.2 & 1.1$^{\dagger}$ \\
\texttt{tmean} & 1.7 & 0.8 & 0.8$^{\dagger}$ & 1.6$^{\dagger}$ & 0.1$^{\dagger}$ & 0.1 & 0.4 & 0.7 & 3.2 & \textbf{1.4} & \textbf{4.3} & 1.8 & \textbf{1.1} & \textbf{3.2} & 0.9 & \textbf{1.9} & 0.5$^{\dagger}$ & 0.0$^{\dagger}$ \\
\texttt{wim\_ethfrac} & 0.6$^{\dagger}$ & 1.5$^{\dagger}$ & 0.2 & 0.2 & \textbf{1.0} & 1.7 & 0.4$^{\dagger}$ & 0.1 & 2.8 & \textbf{2.1} & 3.6 & 1.5 & \textbf{0.5} & 2.6 & 0.6 & 0.4$^{\dagger}$ & 0.5 & 1.2$^{\dagger}$ \\
\bottomrule
\end{tabular}
\end{sidewaystable}

\section{ICE Response Curves for Table 3 Edges}
\label{app:ice}

\begin{figure}[!htbp]
\centering
\includegraphics[width=\textwidth]{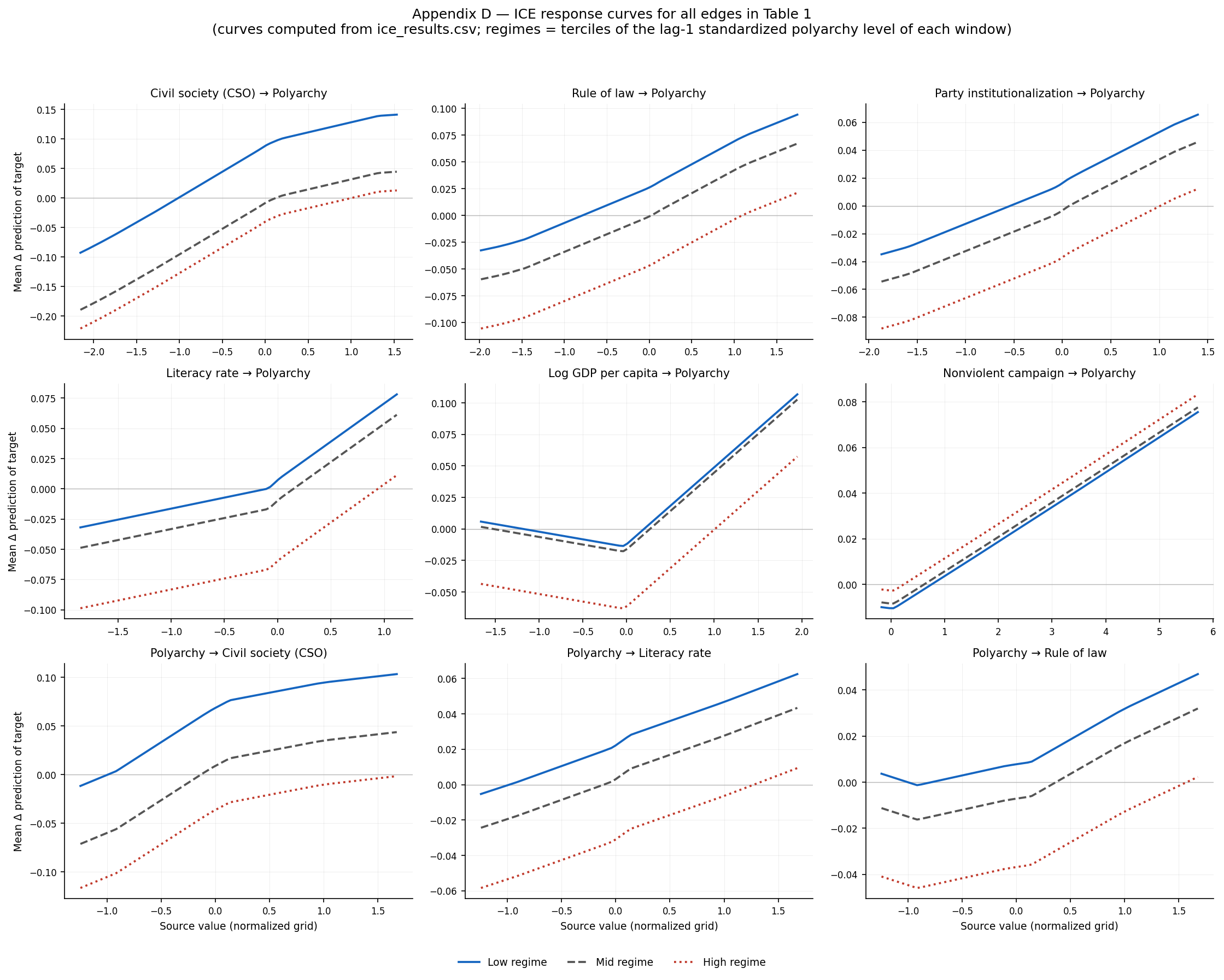}
\caption{ICE response curves for the nine edges in
Table~1 of the main text, computed from the pipeline's ICE output.
Each panel overlays the low (solid blue), mid (dashed grey), and high
(dotted red) democratic regimes, defined as terciles of the lag-1
standardized polyarchy level of the prediction window. Horizontal axis:
source value on the normalized 81-point grid; vertical axis: mean change
in the predicted target. Curves are plotted as estimated, with no
smoothing beyond the model itself.}
\label{fig:ice}
\end{figure}

Figure~\ref{fig:ice} plots the estimated ICE response curves for the
nine edges in Table~3 of the main text, computed directly from the
pipeline's ICE output (81-point grid over each source's empirical
range; regimes are as defined in Section~3.4 of the main text.). Each panel overlays the three democratic regimes, defined as terciles of the lag-1
standardized polyarchy level of the prediction window. The curves
ground the functional-form labels in Table~3 of the main text: the
six edges labeled ``linear $\to$ saturating'' show a near-constant
slope in the low-democracy regime that flattens visibly in the high
regime; the GDP per capita edge changes slope direction within every
regime; and its shape deserves attention because it departs from the
functional form the literature leads one to expect. The source
variable is already logged, and the conventional expectation is that
democracy responds linearly, if at all, to log income. The estimated
curves instead bend within log income itself: the response is
approximately flat in the low- and mid-democracy regimes and mildly
negative in the high regime below roughly the sample average of
logged GDP per capita, and rises approximately linearly in logged
income above that point. The implication is that the logarithmic
transformation alone does not capture the relationship: a threshold
within log income separates a flat-to-negative segment from a linear
one. Threshold formulations of modernization theory anticipate a
development floor of this kind \parencite{huntington1991,
przeworski1997}, and nonlinear re-estimations of the
income--democracy relationship have similarly reported that the
association emerges where linear-in-logs specifications miss it
\parencite{acemoglu2008, benhabib2013};
The classification compresses, yet does not discard the available information. The estimated curves are retained in full and reported as estimated: this appendix plots every relationship in Table~1 of the main text across all three regimes, and any analysis can work from the curves directly. The labels serve the purposes the raw curves cannot. Statements about the system as a whole, such as the share of relationships that are nonlinear in some regime, require a common vocabulary; and theoretical claims become testable only when a curve's shape is named, since ``civil society's effect saturates as democracy consolidates'' can be checked and contradicted in a way that an unlabeled curve cannot \parencite{goldstein2015, molnar2022}. Where a curve resists the taxonomy, the resistance is reported rather than suppressed: two relationships in the application fall outside the four classes and are labeled accordingly.

Curves are plotted exactly as estimated, with no smoothing beyond what the model's own functional form imposes.

\section{Linear Benchmark, Full Results}
\label{app:benchmark}
\emph{Specification.} Pooled OLS and ridge regression were run on the
flattened $(N \cdot K)$-regressor windows, with identical construction and
identical terminal validation split. Edge verdicts from block tests were
that all eight lag coefficients of a source are zero in the target
equation (classical $F$ and HC1 Wald). Ridge penalty was selected from
$\{0.1, 1, 10, 100, 1000\}$ by validation MSE ($\alpha = 1.0$
selected). Cluster-robust inference is not available at this
parameterization (113 clusters against 185 parameters per equation);
both reported tests are likely optimistic under within-country serial
dependence, which is conservative for null verdicts. The forecast comparison also underwrites the necessity criterion
itself: forecast necessity is defined relative to the model class
doing the forecasting, and it carries force only because that class
predicts the system better than the linear alternative given every
specification advantage.

The benchmark fits a pooled linear autoregression on the identical
23-variable panel, 8-lag windows, and terminal validation split used
by the pipeline as described above. Table~\ref{tab:benchmse} reports
forecast accuracy; Table~\ref{tab:benchedges} reports the complete
outcome equation: block tests on all eight lag coefficients of every
source in the polyarchy equation, and the three feedback edges of
Table~1 of the main text. Both $p$-value columns are likely
optimistic under within-country serial dependence, which is
conservative for null verdicts. The NAVAR reference in
Table~\ref{tab:benchmse} is the seed-42 masked validation loss from
the ensemble run of Appendix~\ref{app:ensemble}; the split and
criterion are identical by construction. Full $23 \times 23$ matrices
of block $p$-values (classical and robust) and lag-coefficient sums
for all 396 directed pairs are provided as supplementary files
(\texttt{linear\_granger\_pvalues.csv},
\texttt{linear\_granger\_pvalues\_hc1.csv},
\texttt{linear\_lagsum\_coefs.csv}).

\begin{table}[!htbp]
\centering
\caption{Masked validation MSE, dynamic targets, terminal 10\% split.}
\label{tab:benchmse}
\small
\begin{tabular}{@{}lc@{}}
\toprule
Model & Masked val.\ MSE \\
\midrule
OLS (23 variables, $K=8$)        & 0.2289 \\
Ridge ($\alpha = 1.0$, grid-selected) & 0.2270 \\
NAVAR (seed 42)                  & 0.2032 \\
\bottomrule
\end{tabular}
\end{table}

\begin{table}[!htbp]
\centering
\caption{Linear edge verdicts. Block test that all eight lag
coefficients of the source are zero in the target equation; lag-sum is
the sum of the eight coefficients.}
\label{tab:benchedges}
\small
\begin{tabular}{@{}lccc@{}}
\toprule
Source $\rightarrow$ Polyarchy & $p$ (classical $F$) & $p$ (HC1 Wald) & Lag-sum \\
\midrule
Civil society (CSO)           & $<$0.0001 & 0.0040 & $+0.0140$ \\
Nonviolent campaign           & $<$0.0001 & 0.0001 & $+0.0206$ \\
Literacy rate                 & $<$0.0001 & 0.0021 & $+0.0165$ \\
Party institutionalisation    & 0.0003 & 0.0717 & $+0.0033$ \\
Strict presidentialism        & 0.0006 & 0.1283 & $-0.0015$ \\
Rule of law                   & 0.0021 & 0.0538 & $+0.0086$ \\
Protestant population         & 0.0215 & 0.1021 & $+0.0004$ \\
National econ.\ growth        & 0.0316 & 0.3872 & $+0.0036$ \\
Anti-system movement          & 0.0329 & 0.2775 & $-0.0060$ \\
Majoritarian electoral system & 0.0365 & 0.1747 & $-0.0073$ \\
Left-wing movement            & 0.0588 & 0.6376 & $-0.0045$ \\
Right-wing movement           & 0.0819 & 0.3402 & $-0.0046$ \\
European population           & 0.0836 & 0.5242 & $+0.0067$ \\
Global econ.\ growth          & 0.0873 & 0.2042 & $-0.0029$ \\
Log GDP per capita            & 0.1632 & 0.5906 & $+0.0020$ \\
Violent campaign              & 0.2038 & 0.0777 & $-0.0046$ \\
Mean temperature              & 0.3691 & 0.2985 & $+0.0047$ \\
Ongoing campaign              & 0.3863 & 0.1995 & $+0.0084$ \\
Agricultural income share     & 0.4656 & 0.3953 & $-0.0052$ \\
Ethnic fractionalization      & 0.8755 & 0.8657 & $-0.0006$ \\
Harbor distance               & 0.8909 & 0.5116 & $+0.0019$ \\
Resource dependence           & 0.9915 & 0.8867 & $-0.0040$ \\
\midrule
\multicolumn{4}{@{}l}{\emph{Feedback direction}} \\
Polyarchy $\rightarrow$ Civil society (CSO) & 0.1979 & 0.5264 & $+0.0133$ \\
Polyarchy $\rightarrow$ Literacy rate       & 0.0001 & 0.1464 & $-0.0012$ \\
Polyarchy $\rightarrow$ Rule of law         & 0.3934 & 0.5613 & $+0.0085$ \\
\bottomrule
\end{tabular}
\end{table}

\section{Seed-Ensemble Robustness}
\label{app:ensemble}

Neural estimation of the pipeline is sensitive to the random seed
through weight initialization, batch shuffling, and dropout. All point
values in the main text come from a single documented seed (42). This
appendix reports a five-seed ensemble (seeds 42, 7, 123, 2024, 31415)
in which Stages~2 through~3b were re-estimated per seed with data,
hyperparameters, lag windows, and the holdout construction held fixed,
so that variation isolates the stochastic components of training.

\subsection{Ensemble-level quantities}

Forecast-necessary edge counts are stable across seeds (156--167 of 396
candidates). The automatic GMM threshold varies from
$5.5 \times 10^{-4}$ to $9.4 \times 10^{-4}$, and the resulting
dominant-necessary graph contains between 59 and 84 edges; the seed-42
graph reported in the main text is the largest in the ensemble.
Forty-nine edges are retained under all five seeds, 71 under a
majority, and 92 under at least one. We refer to the 49-edge
intersection as the \emph{stable core}. The main text's aggregate
nonlinearity finding is unchanged on the core: 42 of 49 core edges
(86\%) are nonlinear in at least one democratic regime, against 73 of
84 (87\%) on the seed-42 graph.

\subsection{Headline edges}

Table~\ref{tab:ensemble} reports retention frequency and, conditional
on retention, ensemble score statistics for every edge in
Table~1 of the main text. Following our reporting convention for
seed-sensitive quantities, retention frequency is the primary quantity
and score ranges are reported as min--max membership rather than point
matches.

\begin{table}[!htbp]
\centering
\caption{Seed-ensemble robustness of headline edges (5 seeds). Score
statistics are computed over seeds in which the edge is retained.}
\label{tab:ensemble}
\small
\begin{tabular}{@{}lccc@{}}
\toprule
Edge & Retention & Score mean (sd) & Score range \\
\midrule
Civil society (CSO) $\rightarrow$ Polyarchy        & 5/5 & 0.072 (0.001) & [0.070, 0.073] \\
Rule of law $\rightarrow$ Polyarchy                & 5/5 & 0.041 (0.001) & [0.039, 0.042] \\
Party institutionalization $\rightarrow$ Polyarchy & 5/5 & 0.031 (0.001) & [0.030, 0.033] \\
Literacy rate $\rightarrow$ Polyarchy              & 4/5 & 0.039 (0.001) & [0.038, 0.041] \\
Log GDP per capita $\rightarrow$ Polyarchy         & 4/5 & 0.034 (0.002) & [0.033, 0.037] \\
Nonviolent campaign $\rightarrow$ Polyarchy        & 3/5 & 0.031 (0.001) & [0.030, 0.031] \\
Polyarchy $\rightarrow$ Civil society (CSO)        & 5/5 & 0.047 (0.002) & [0.045, 0.049] \\
Polyarchy $\rightarrow$ Literacy rate              & 5/5 & 0.024 (0.001) & [0.023, 0.026] \\
Polyarchy $\rightarrow$ Rule of law                & 5/5 & 0.023 (0.001) & [0.022, 0.025] \\
Agricultural income $\rightarrow$ Polyarchy        & 0/5 & --- & --- \\
\bottomrule
\end{tabular}
\end{table}

\subsection{Drop mechanisms for non-universal edges}

The three edges below universal retention fail in two distinct ways,
with different interpretations. The literacy and GDP edges are
\emph{threshold-clipped} in the single seed (7) that omits them: their
mean loss differentials remain positive ($1.5 \times 10^{-4}$ and
$4.2 \times 10^{-4}$) but fall below that seed's automatic threshold
($6.7 \times 10^{-4}$). The necessity signal weakens under one seed;
it does not reverse. The nonviolent-campaign edge, by contrast,
\emph{fails the DM test itself} in the two seeds that omit it, despite
loss differentials above the threshold in both---its forecast
necessity, not merely its dominance, is seed-dependent. This is the
basis for the ``suggestive'' framing in Section~5 of the main text.

\subsection{What is and is not seed-stable}

Stable under all five seeds: the three protective-belt edges into
polyarchy; eight feedback edges from polyarchy, including all three
reported in Table~1 of the main text; the absence of a direct
agricultural-income edge (zero of five seeds), alongside universal
retention of four of its six indirect paths; and the predominance of
nonlinear functional forms. Retained in four of five seeds: the two
overturned modernization nulls. Retained in three: the
nonviolent-campaign edge. The replication (RQ1) and augmentation (RQ2)
findings are therefore seed-robust in full; the reinterpretation
findings (RQ3) hold at the stated frequencies. Per-seed output files
and the ensemble driver are included in the replication package.

\end{document}